\documentclass[aps,prb,twocolumn,showpacs]{revtex4-1}
\usepackage{amsmath,amssymb}
\usepackage{color,framed}
\usepackage[pdftex]{hyperref,graphicx}
\usepackage{xspace}
\usepackage{hypcap}
\usepackage{dsfont}
\usepackage{listliketab}
\usepackage{multirow}
\usepackage{tabularx}
\usepackage{cleveref}

\crefname{pluralequation}{eqs.}{eqs.}
\Crefname{pluralequation}{Eqs.}{Eqs.}

\crefformat{equation}{#2~(#1)#3}
\Crefformat{equation}{#2~(#1)#3}

\hypersetup{colorlinks = true, urlcolor = blue, linkcolor = blue, citecolor = blue}

\begin{document}
\title{Transport in superconductor--normal metal--superconductor tunneling structures: Spinful $p$-wave and spin-orbit-coupled topological wires}

\author{F.~Setiawan}\email{setiawan@umd.edu}
\author{William~S.~Cole}
\author{Jay~D.~Sau}
\author{S.~Das~Sarma}
\affiliation{Condensed Matter Theory Center, Station Q Maryland, and Joint Quantum Institute,
Department of Physics, University of Maryland, College Park, Maryland 20742, USA}
\date{\today}

\begin{abstract}

We theoretically study transport properties of voltage-biased one-dimensional superconductor--normal metal--superconductor tunnel junctions with arbitrary junction transparency where the superconductors can have trivial or nontrivial topology. Motivated by recent experimental efforts on Majorana properties of superconductor-semiconductor hybrid systems, we consider two explicit models for topological superconductors: (i) spinful $p$-wave, and (ii) spin-split spin-orbit-coupled $s$-wave. We provide a comprehensive analysis of the zero-temperature dc current $I$ and differential conductance $dI/dV$ of voltage-biased junctions with or without Majorana zero modes (MZMs). The presence of an MZM necessarily gives rise to two tunneling conductance peaks at voltages $eV = \pm \Delta_{\mathrm{lead}}$, i.e., the voltage at which the superconducting gap edge of the lead aligns with the MZM. We find that the MZM conductance peak probed by a superconducting lead \emph{without} a BCS singularity has a nonuniversal value which decreases with decreasing junction transparency. This is in contrast to the MZM tunneling conductance measured by a superconducting lead \emph{with} a BCS singularity, where the conductance peak in the tunneling limit takes the quantized value $G_M = (4-\pi)2e^2/h$ independent of the junction transparency.  We also discuss the ``subharmonic gap structure", a consequence of multiple Andreev reflections, in the presence and absence of MZMs. Finally, we show that for finite-energy Andreev bound states (ABSs), the conductance peaks shift away from the gap bias voltage $eV = \pm \Delta_{\mathrm{lead}}$ to a larger value set by the ABSs energy. Our work should have important implications for the extensive current experimental efforts toward creating topological superconductivity and MZMs in semiconductor nanowires proximity coupled to ordinary $s$-wave superconductors.
\end{abstract}


\maketitle

\newcommand{\beq}{\begin{equation}}
\newcommand{\eeq}{\end{equation}}
\newcommand{\ba}{\begin{align}}
\newcommand{\ea}{\end{align}}
\newcommand{\nl}{\nonumber\\}
\newcommand{\ssc}{$s$SC\xspace}
\newcommand{\psc}{$p$SC\xspace}
\newcommand{\bmat}{\begin{matrix}}
\newcommand{\emat}{\end{matrix}}
\newcommand{\sgn}{\operatorname{sgn}}
\newcommand{\jtle}{\widetilde{j}_{NL,\nu}^e}
\newcommand{\jtlh}{\widetilde{j}_{NL,\nu}^h}
\newcommand{\jtre}{\widetilde{j}_{NR,\nu}^e}
\newcommand{\jtrh}{\widetilde{j}_{NR,\nu}^h}
\newcommand{\iinpl}{\widetilde{j}^{\mathrm{tot}}_{NL,\nu}}
\newcommand{\iinpr}{\widetilde{j}^{\mathrm{tot}}_{NR,\nu}}
\newcommand{\dmin}{\Delta_{\mathrm{min}}\xspace}
\newcommand{\dminl}{\Delta_{\mathrm{eff,L}}\xspace}
\newcommand{\dminr}{\Delta_{\mathrm{eff,R}}\xspace}
\newcommand{\dtsoc}{\Delta^{\mathrm{topo}}_{\mathrm{SOCSW}}\xspace}
\newcommand{\dntsoc}{\Delta^{\mathrm{nontopo}}_{\mathrm{SOCSW}}\xspace}

\section{Introduction}
In recent years there has been great interest in realizing topological superconductors that support Majorana zero modes (MZMs) at boundaries or defects~\cite{Alicea12,Beenakker,Leijnse,Tudor,Franz,Sarma,beenakker16}. This is driven mainly by the prospect of using MZMs as the building blocks for a fault-tolerant topological quantum computer~\cite{kitaev03,Nayak}. The simplest model of a topological superconductor hosting MZMs is the one-dimensional (1D) spinless $p$-wave superconductor as originally envisioned by Kitaev~\cite{Kitaev}. Since electrons carry a spin degree of freedom, intrinsic spinless $p$-wave pairing is apparently uncommon in nature. However, it can be effectively realized in spinful systems by a combination of spin-orbit coupling and explicitly lifting the Kramer's degeneracy of the electrons (e.g., by Zeeman spin splitting through an applied magnetic field). This idea has lead to a number of proposals for realizing topological superconductor in various hybrid structures with conventional $s$-wave superconductors~\cite{Fu08,roman,oreg,jay09,jay10,alicea,Fu09,Mi13,Choy11,Duckheim11,Chung11,Mao12,Sau12,Kim14,Brydon15,Hoi15,Perg}. There are, however, significant differences between a spinless $p$-wave superconductor and a spin-split $s$-wave superconductor with spin-orbit coupling although they both can have localized MZMs at the ends. One way of realizing a spinful 1D topological superconductor is by proximity-inducing superconductivity in a spin-orbit-coupled semiconducting nanowire in a magnetic field~\cite{roman,oreg,jay10,alicea}. In this setup, the system can be tuned from a topologically trivial to a nontrivial regime by raising the Zeeman field above a certain critical value where the system undergoes a topological quantum phase transition with the effective induced superconductivity in the nanowire changing from an $s$-wave (trivial) character to a $p$-wave (topological) character. As the MZM exists as a zero-energy edge mode in the topological superconductor, tunneling conductance spectroscopy provides a simple way of detecting the MZM. In a normal metal--superconductor (NS) junction, the MZM mediates a perfect Andreev reflection at zero energy, which in turn gives rise to a quantized $2e^2/h$ zero-bias conductance value~\cite{Sengupta01,Law09,Flensberg10,Wimmer11,Setiawan15}, as long as the two MZMs at the wire ends are far from each other with exponentially small overlap between the MZM wave functions (the so-called ``topologically protected regime"). This quantized conductance is robust against changes to the junction transparency. Several experimental groups have observed the appearance of zero-bias tunneling conductance peaks in the semiconductor--superconductor heterostructure as the Zeeman field is raised beyond a certain value, which indeed indicates the existence of zero-energy states~\cite{Mourik,deng12,Das,Churchill,Finck,Zhang16,Deng16,Frolov}. Nevertheless, the observed zero-bias conductance is substantially less than the MZM canonical quantized conductance value. A plausible source for this discrepancy is thermal broadening in the normal-metal lead, which reduces the zero-bias conductance value and widens the peak~\cite{Sengupta01,Lin12,Wimmer11,Reeg}, although other possibilities such as dissipation and MZM overlap may also be responsible~\cite{Lin12,Nag,Chunxiao}. The ubiquitous absence of the predicted quantized zero-bias conductance peak in topological NS junction (in spite of there often being a weak zero-bias conductance peak) is the central quandary in this subject, making it unclear whether 1D topological superconductivity with localized MZMs has indeed been realized experimentally or not.

To mitigate the effect of thermal broadening, one could use a superconducting lead instead of a normal lead in probing the MZM tunneling conductance~\cite{Peng15,Yeyati,seti2016,Denis}. In a superconducting lead, thermal quasiparticle excitations are exponentially suppressed by the superconducting gap $\sim\mathrm{exp}(-\Delta_{\mathrm{lead}}/T)$, which in turn suppresses the broadening effect. Peng \textit{et al.}~\cite{Peng15} found that for a conventional $s$-wave superconducting lead, the MZM conductance measured in the tunneling limit (or small junction transparency) appears as two symmetric peaks with a quantized value $G_M = (4-\pi)2e^2/h$ at the gap-bias voltage $eV = \pm \Delta_{\mathrm{lead}}$, i.e., when the BCS singularity of the probe lead aligns with the MZM. The quantized value $G_M$ is the conductance due to a single Andreev reflection from the MZM. In Ref.~\cite{seti2016}, it was shown that in the presence of multiple Andreev reflections (MAR), which are generically present when the junction transparency is \emph{not} small, the conductance at the voltage $eV = \pm \Delta_{\mathrm{lead}}$ is no longer quantized at $G_M$. This indicates that unlike the universally quantized $2e^2/h$ zero-bias conductance value for a normal metal--topological superconductor junction, the quantized value $G_M$ of the MZM tunneling conductance for a superconductor--normal metal--superconductor (SNS) junction is not a topologically protected robust quantity with the conductance value being dependent on the details and thus making it difficult to identify MZMs using SNS tunneling spectroscopy. These results prompt further exploration of transport properties of various SNS junctions involving different models of topological superconductors where signatures of MZM can be fully investigated and characterized. This is the goal of the current work where we provide comprehensive details on the tunneling transport properties of SNS junctions involving topological superconductors in order to guide future experimental work in the subject.

In this paper, we calculate the dc current-voltage ($I$-$V$) relation and corresponding differential conductance ($G = dI/dV$) of 1D SNS junctions, invoking two models for topological superconductors, i.e., the spinful $p$-wave superconductor (\psc) and the spin-orbit-coupled $s$-wave superconducting wire (SOCSW). Specifically, we consider several possible combinations for the junction, where each superconductor can be either topologically trivial or nontrivial. We find that unlike the case of $s$-wave superconducting probe lead with BCS singularity (where $\sum_{\sigma = \uparrow,\downarrow}|u_{\sigma}|^2 = \sum_{\sigma = \uparrow,\downarrow}|v_{\sigma}|^2$ at the gap edge with $u$ and $v$ being the electron and hole component of the superconducting wavefunction at the gap edge), the MZM tunneling conductance measured using a superconducting lead without a BCS singularity has a nonuniversal value, which decreases with decreasing junction transparencies. Our detailed theoretical and numerical results for the transport properties of various types of SNS junctions should be a useful guide for future experimental work on the tunneling spectroscopy of topological SNS junctions.

The paper is organized as follows. In Sec.~\ref{sec2}, we present a general scattering matrix formalism, which can be used to calculate the transport properties of a general SNS junction. In Sec.~\ref{sec:SGS}, we discuss the subharmonic gap structure (SGS). In the following sections, we study in detail the transport in SNS junctions involving \psc (Sec.~\ref{sec3}) and SOCSW (Sec.~\ref{sec4}) and compare the dc current and conductance of junctions with and without MZMs. In Sec.~\ref{sec5}, we discuss the conductance due to Andreev bound states (ABSs) in the SOCSW model in order to distinguish between MZM and ABS signatures in the tunneling experiment. Finally, we give the conclusion in Sec.~\ref{sec6}.

\section{Scattering Matrix Formalism}\label{sec2}
\begin{figure}[h]
\capstart
\begin{center}\label{Fig1}
\includegraphics[width=\linewidth]{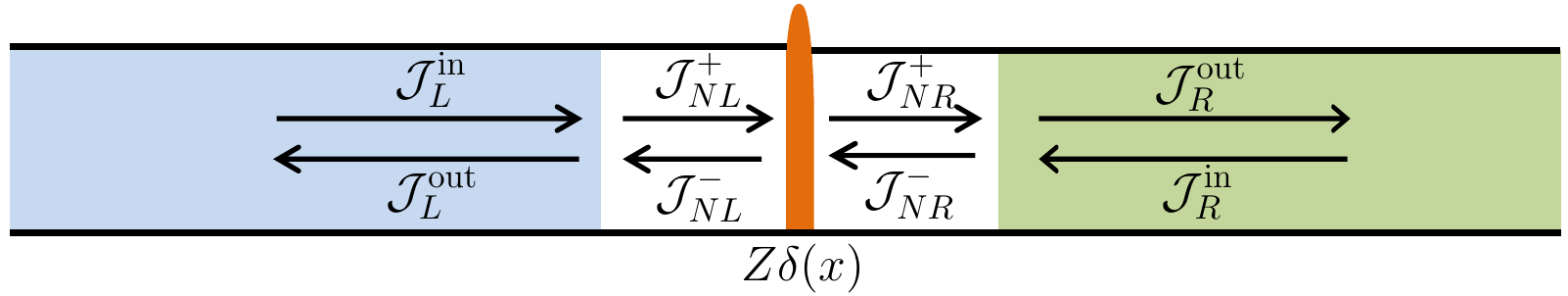}
\end{center}
\caption{(Color online) Schematic diagram of a superconductor--normal metal--superconductor (SNS) junction with a delta-function potential barrier of strength $Z$.} 
\end{figure}

We begin by modeling the SNS junction by two semi-infinite superconducting regions connected by a normal region with a delta-function barrier of strength $Z$, as shown in Fig.~\ref{Fig1}. The normal region is assumed to be infinitesimally short with large chemical potential such that the propagating modes in this region have constant group velocity independent of energy. Quasiparticles can be injected from the left or right superconducting lead which become electrons or holes (depending on their energy) when they enter the normal region. Due to the voltage bias, these electrons (holes) will then gain (lose) an energy $eV$ as they are accelerated from the left (right) to the right (left). As a result, after each Andreev reflection at an NS interface, an incoming electron with an energy $E$ will be reflected as a hole back into the same region with an energy $E+2eV$. The quasiparticle reflects repeatedly inside the normal region until it gains enough energy to be transmitted into the superconductors. This mechanism is termed ``multiple Andreev reflections" (MAR)~\cite{Averin,octavio,KBT}. 

The scattering processes in the SNS junction can be split among three spatial regions: (i) the left NS interface, (ii) tunnel barrier, and (iii) right NS interface. We express these processes in terms of the scattering matrices as 
\begin{subequations}\label{eq:matchingeq}\label[pluralequation]{eq:matchingeq}
\begin{align}
\left( \bmat
\mathcal{J}_{L,\nu}^{\mathrm{out}} (E_n) \\
\mathcal{J}_{NL,\nu}^{+} (E_n)
\emat\right) &= S_L(E_n) \left(\bmat \mathcal{J}_{L,\nu}^{\mathrm{in}} (E_n)\delta_{n0}\delta_{\nu,\rightarrow} \\\mathcal{J}_{NL,\nu}^{-} (E_n)\emat  \right)\label{subeq1}, \\
\left( \bmat
\mathcal{J}_{NL,\nu}^{-} (E_n) \\
\mathcal{J}_{NR,\nu}^{+} (E_n) 
\emat\right) &= \sum_{n'} S_N(E_n,E_{n'}) \left( \bmat \mathcal{J}_{NL,\nu}^{+} (E_{n'}) \\ \mathcal{J}_{NR,\nu}^{-} (E_{n'}) \emat\right)\label{subeq2},\\
\left( \bmat
\mathcal{J}_{R,\nu}^{\mathrm{out}} (E_n)\\
\mathcal{J}_{NR,\nu}^{-} (E_n) 
\emat\right) &= S_R(E_n) \left(\bmat \mathcal{J}_{R,\nu}^{ \mathrm{in}} (E_n)\delta_{n0}\delta_{\nu,\leftarrow} \\ \mathcal{J}_{NR,\nu}^{+} (E_n) \emat  \right)\label{subeq3},
\end{align}
\end{subequations} 
where $E_n = E+ neV$ are the energies of propagating modes with $n$ being an integer, $\mathcal{J}_{\ell,\nu}^\rho = (j_{\ell,\nu}^{e,\uparrow,\rho},j_{\ell,\nu}^{e,\downarrow,\rho},j_{\ell,\nu}^{h,\uparrow,\rho},j_{\ell,\nu}^{h,\downarrow,\rho})^{\mathrm{T}}$ is the current amplitude vector for region $\ell = L$ (left superconductor), $NL$ (normal region to the left of the tunnel barrier), $NR$ (normal region to the right of the tunnel barrier), and $R$ (right superconductor) with $\rho = +/-$ and $\rho = \mathrm{in}/\mathrm{out}$ being the right/left-moving modes and incoming/outgoing modes indices, respectively, and $\nu = \rightleftarrows$ denoting whether the incoming quasiparticle is from the left or right superconductor. We note that the scattering matrix formalism presented above is completely general and can be utilized to study the transport properties of any kind of SNS junctions with arbitrary topological properties (including trivial superconductors). Moreover, it can be easily interfaced with the numerical transport package Kwant~\cite{kwant}, which can be used to calculate the scattering matrices of the left ($S_L$) and right ($S_R$) NS interfaces~\cite{seti2016,BTK}. For details on our numerical simulations, see the Appendix. 

The scattering matrix $S_N(E_n,E_n')$ in Eq.~\cref{subeq2} incorporates the scattering processes at the tunnel barrier and the increase (decrease) of the electron (hole) energy by $eV$ each time the electron (hole) passes from the left to the right. In terms of the electron ($S_{N}^e$) and hole ($S_{N}^h$) component, it can be written as 
\begin{align}
S_N(E_n,E_{n'}) \;=\; &S_{N}^e(E_n,E_{n'})\otimes \sigma_0 \otimes \tau_+  \nonumber \\ & + S_{N}^h(E_n,E_{n'}) \otimes \sigma_0\otimes\tau_-,
\end{align}
where $\sigma_0$ is the identity matrix in the spin subspace, $\tau_\pm = \tau_x \pm i\tau_y$ are the Pauli matrices in the particle-hole subspace. The scattering matrices $S_{N}^e$ and $S_{N}^h$ are given by
\begin{align}
S_{N}^e(E_n,E_{n'}) &= \left(\bmat r \delta_{n,n'} & t \delta_{n,n'+1} \\ t \delta_{n,n'-1} & r \delta_{n,n'} \emat \right), \nl
S_{N}^h(E_n,E_{n'}) &= \left(\bmat r^* \delta_{n,n'} & t^* \delta_{n,n'-1} \\ t^* \delta_{n,n'+1} & r^* \delta_{n,n'} \emat \right),
\end{align}
where $r = -iZ/(1+iZ)$ and $t = 1/(1+iZ)$ are the reflection and transmission coefficients, respectively, with the amplitudes depending on the delta-function barrier strength $Z$. Note that $Z$ should be considered an effective barrier strength determining the interface transparency represented by a delta-function potential which is an unknown parameter in the theory (as in the well-known Blonder-Tinkham-Klapwijk (BTK) formalism~\cite{BTK}). In principle, $Z$ can be calculated from first principles if all information about the interface is available, but in practice $Z$ should be determined by comparing theory and experiment. We change the junction transparency in the simulation by tuning $Z$. Since sharp changes of parameters across the junction, such as mismatch in the Fermi level, spin-orbit coupling, $p$-wave pairing potential etc., also effectively create barriers for the current, we use a parameter-independent quantity $G_N$ to characterize the junction transparency, where $G_N$ is the normalized conductance of the SNS junction at high voltages (in the unit of $G_0 = e^2/h$), which is the conductance of the corresponding normal-normal (NN) junction. We note that $G_N$, which subsumes $Z$ and other possible unknown microscopic parameters, can be directly measured experimentally allowing experiment and theory to be quantitatively compared at arbitrary voltages.  We refer to $G_N$ as the ``junction transparency" in the rest of this paper since it denotes the conductance for the corresponding NN junction.  Note that $G_N=2$ or $G_N=1$ denotes perfect transparency (corresponding to $Z$=0) depending on the specific tunnel junction one is considering.

Solving the coupled linear equations [Eq.~\eqref{eq:matchingeq}], we obtain the current amplitudes $\mathcal{J}_{\ell,\nu}^\rho$. The total current can be calculated by adding up the contribution from the left- and right-moving modes of the electrons and holes for the incoming quasiparticles from the left and right superconductors, i.e.,
\begin{align}\label{eq:curr}
I_{\nu}(V) = \frac{2 e}{h}\int_{-\infty}^{0} dE \mathrm{Tr}\left(\sum_{n} \rho_z \tau_z J_{NL,\nu}(E_n) J_{NL,\nu}^\dagger(E_n)\right),
\end{align}
where 
\begin{align}
J_{NL,\nu} =& \left(
j_{NL,\nu}^{e,\uparrow,+}, j_{NL,\nu}^{e,\downarrow,+}, j_{NL,\nu}^{h,\uparrow,+}, j_{NL,\nu}^{h,\downarrow,+}, \right. \nonumber \\
&\left. \hspace{2 cm} j_{NL,\nu}^{e,\uparrow,-}, j^{e,\downarrow,-}_{NL,\nu}, j_{NL,\nu}^{h,\uparrow,-}, j_{NL,\nu}^{h,\downarrow,-}\right)^{\mathrm{T}}
\end{align}
is the current amplitude vector in the normal region to the left of the barrier. It was proven in Ref.~\cite{seti2016} that the current is nonnegative for positive $V$. The differential conductance ($G = dI/dV$) can be computed by directly differentiating the current $I$ with respect to the voltage $V$. In general, we observe that the differential conductance is particle-hole \emph{asymmetric} except for sufficiently small transparencies.

In this paper, we apply this scattering matrix formalism to calculate the zero-temperature dc current and conductance of junctions composed of spinful \psc (Sec.~\ref{sec3}) and SOCSW (Sec.~\ref{sec4}), considering scenarios where none, one, or both of the superconductors are topologically nontrivial.

\section{Subharmonic gap structure}\label{sec:SGS}
In general, for SNS junctions with asymmetric gap ($\Delta_{L} \neq \Delta_{R}$), where $\Delta_{L,R}$ are the superconducting gap of the left and right superconductors, when the junction transparency is not small, there will be nonanalyticities in the $I$-$V$ curve or conductance~\cite{KBT,octavio,Averin} at specific voltages, which is termed the ``subharmonic gap structure" (SGS). The sharp change in the conductance happens at voltages at which there is a change in the number of Andreev reflections required to transfer charge from the occupied to the empty band. For incoming quasiparticles from the left superconductor, this number of Andreev reflections changes when 
\begin{equation}\label{eq:SGSvolt1}
e|V|= \frac{\Delta_{L}}{n}, \hspace{0.5 cm} n \geq 1,
\end{equation}
and 
\begin{equation}\label{eq:SGSvolt2}
e|V| = \frac{\Delta_{L} + \Delta_{R}}{2n-1}, \hspace{0.5 cm} 1 \leq n \leq \frac{\Delta_{R}}{\Delta_{R} - \Delta_{L}}, 
\end{equation}
while for incoming quasiparticles from the right superconductor, the change happens at voltages given in Eq.~\eqref{eq:SGSvolt2} and
\begin{equation}\label{eq:SGSvolt3}
e|V| = \frac{\Delta_{R}}{n}, \hspace{0.5 cm} 1 \leq n \leq \frac{\Delta_{R}}{\Delta_{R} - \Delta_{L}}.
\end{equation}
Without loss of generality, in the above we assume $\Delta_{R} > \Delta_{L}$. The range of $n$ in Eqs.~\eqref{eq:SGSvolt1}-\eqref{eq:SGSvolt3} gives the voltage range for ``strong" SGS where all Andreev reflections happen inside the superconducting gap. The SGS that occurs outside this range of $n$ is termed ``weak" SGS because the Andreev reflections that happen outside the gap have, in general, small amplitude.  

The SGS (including both weak and strong) happens at the voltages given in Table~\ref{table2} (Refs.~\cite{Hurd,Bagwell}), where $\Delta_{L,R}$ can be any of the gap values in the left and right superconductors when there are multiple superconducting gaps. In general, the SGS is not apparent for near-perfect transparency junction and becomes sharper in the intermediate range of transparencies. Decreasing the transparency further into the tunneling limit will diminish the SGS at small voltages.
 
\begin{table}
\caption{Voltages at which the subharmonic gap structure appears for an asymmetric SNS junction.}\label{table2}
  \begin{tabular*}{\linewidth}{@{\extracolsep{\fill}}lc}
    \hline
    SGS voltage $e|V|$ & Range of $n$  \\ \hline
    $\Delta_{L}/n$ & $n \geq 1$ \\ 
    $(\Delta_{L}+\Delta_{R})/(2n-1)$ & $n \geq 1$  \\
		$\Delta_{R}/n$ & $n \geq 1$  \\
    \hline
  \end{tabular*}
\end{table}

\section{spinful $p$-wave superconductor junctions}\label{sec3} 

\begin{figure}
\capstart
\begin{center}\label{Fig2}
\includegraphics[width=\linewidth]{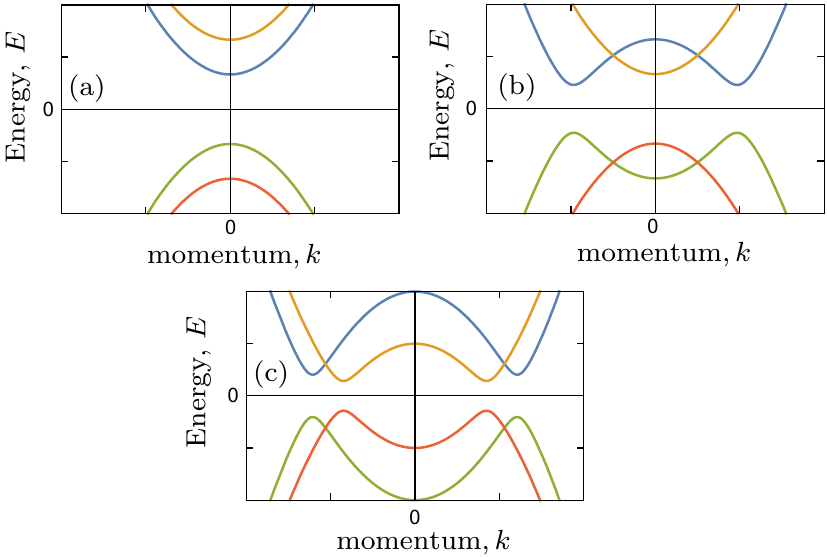}
\end{center}
\caption{(Color online) Energy spectrum of a spinful \psc for different parameter regimes: (a) $|V_Z| < |\mu_p|$ and $\mu_p < 0$ ($p_0$-SC with no topological channel), (b) $|V_Z| > \mu_p$ and $\mu_p > 0$ ($p_1$-SC with one topological channel), (c) $|V_Z| < \mu_p$ and $\mu_p > 0$ ($p_2$-SC with two topological channels). } 
\end{figure}

In this section, we consider junctions between an $s$-wave superconductor (\ssc) and a \psc as well as junctions between two $p$SCs, where the $p$SC can be topologically trivial or nontrivial, depending on the strength of the chemical potential and Zeeman field (i.e., the system is below or above the topological quantum phase transition). Within the Bogoliubov-de Gennes (BdG) formalism, we can write the Hamiltonian of the system as
\begin{equation}\label{eq:Hamilton}
H_j(x) = \frac{1}{2} \int dx \Psi_j^\dagger(x) \mathcal{H}_j \Psi_j(x),
\end{equation}
where $\Psi_j(x) = \left( \psi_{j\uparrow}(x), \psi_{j\downarrow}(x), \psi_{j\downarrow}^\dagger(x),-\psi_{j\downarrow}^\dagger(x) \right)^\mathrm{T}$ are Nambu spinors, and $\psi^\dagger_{j\sigma}(x)$ and $\psi_{j\sigma}(x)$ are the creation and annihilation operators of an electron of spin $\sigma$ for the superconductor of type $j = s,p$ ($s$-wave or $p$-wave). The BdG Hamiltonians for the \ssc and \psc are given by
\begin{subequations}
\begin{align}
\mathcal{H}_s &= \left(-\frac{\hbar^2\partial_x^2}{2 m} - \mu_s\right) \tau_z + \Delta_s \tau_x, \\
\mathcal{H}_p &= \left(-\frac{\hbar^2\partial_x^2}{2 m} - \mu_p\right) \tau_z + V_Z\sigma_z - i\Delta_p \partial_x \tau_x \sigma_x,
\end{align}
\end{subequations}
respectively. Here, $m$ is the electron effective mass (for the numerical simulations done in this paper, we set $m = 0.015m_e$, which corresponds approximately  to InSb nanowires~\cite{Mourik}, where $m_e$ is the bare electron mass), $\mu_s$ and $\mu_p$ are the chemical potentials of the \ssc and \psc, $V_Z$ is the Zeeman field, $\Delta_s$ and $\Delta_{p}$ are the \ssc and \psc pairing potentials, and $\tau_{x,y,z}$ ($\sigma_{x,y,z}$) are Pauli matrices acting in the particle-hole (spin) subspace. The effective chemical potential in each spin channel of the \psc  ($\mu_p \pm V_Z$) determines whether that channel is topological or not. The channel is topological if its chemical potential is positive; otherwise, it is nontopological~\cite{Kitaev,Read}. The spinful \psc can have zero, one, or two topological channels, depending on the values of $V_Z$ and $\mu_p$, i.e.,
\storestyleof{itemize}
\begin{listliketab}
    \begin{tabular}{Lll}
		(a) & $|V_Z| < |\mu_p|$ and $\mu_p < 0$, & no topological channel, \\
		(b) & $|V_Z| > \mu_p$ and $\mu_p > 0$, & one topological channel, \\
		(c) & $|V_Z| < \mu_p$ and $\mu_p > 0$, & two topological channels.
		\end{tabular}
\end{listliketab}
Throughout this paper, we denote the \psc in these three different regimes as $p_i$, where $i=0,1,2,$ refers to the number of topological channels in the \psc.  Since the spinful \psc is essentially made up of two uncoupled spinless $p$SCs, the spectrum of the spinful \psc then consists of the spectrum of two spinless $p$SCs~\cite{Setiawan15} with effective chemical potential $\mu_p \pm V_Z$ as shown in Fig.~\ref{Fig2}. In the following we will denote the smallest gap in the spectrum of the $p_i$-SC by $\Delta_{p_i}$.

\subsection{$sNp_0$ junction}

\begin{figure}[h]
\capstart
\begin{center}\label{Fig3}
\includegraphics[width=\linewidth]{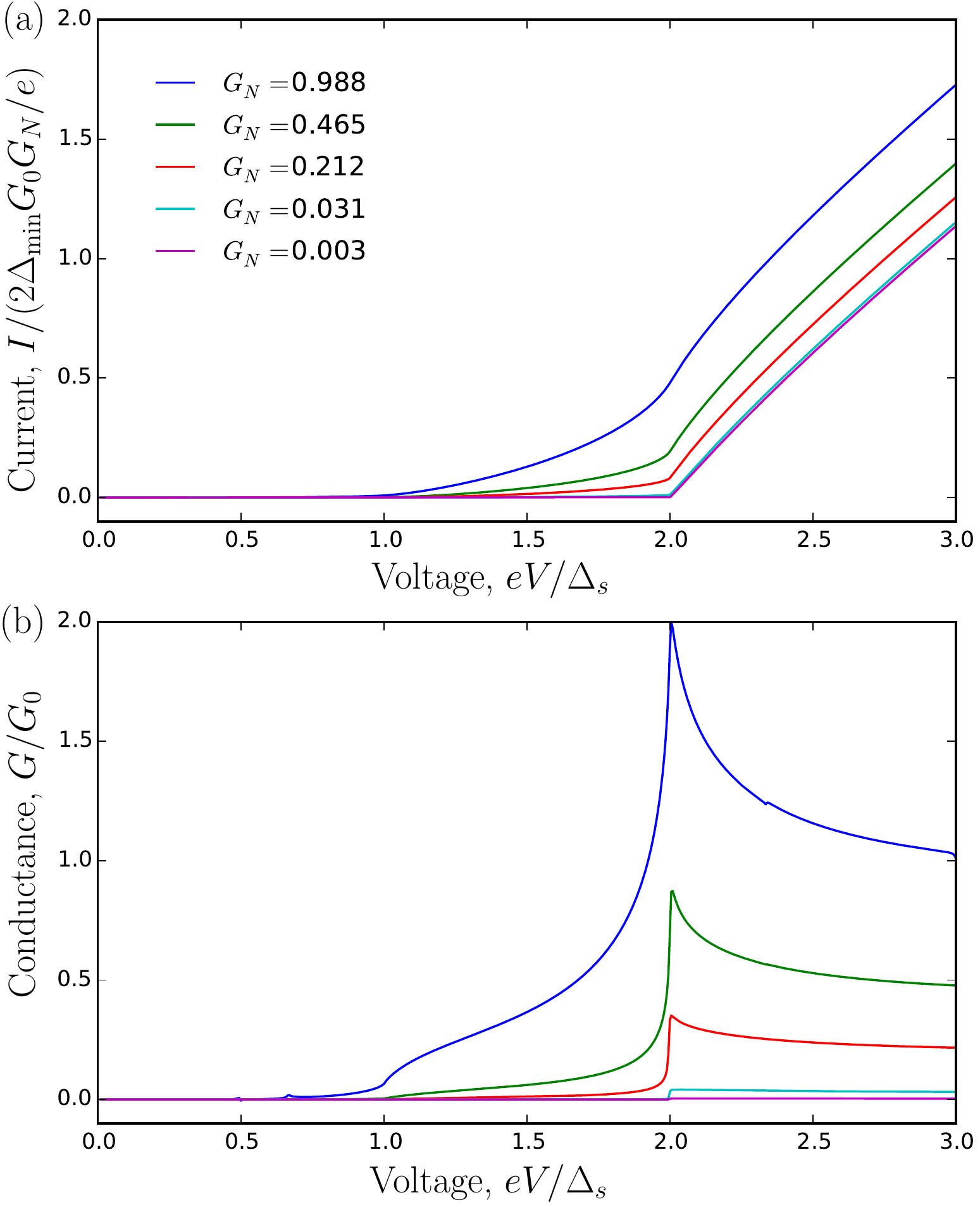}
\end{center}
\caption{(Color online) Plots of (a) dc current $I$ and (b) normalized differential conductance $G/G_0$ versus bias voltage $V$ for an $sNp_0$ junction with various values of transparencies $G_N$. The parameters used for the $s$SC are $\mu_{s} = 20$ K and $\Delta_{s} = 0.01$ K. The parameters used for the $p_0$-SC are $\mu_{p} = -0.01$ K, $V_{Z}= 0$ K, $\Delta_{p} = 0.2$ eV\AA,  where the smallest gap is $\Delta_{p_0} = 0.01$ K. The smallest gap in the junction is $\Delta_{\mathrm{min}}$ = 0.01 K.} 
\end{figure}

We begin by considering the $s$-wave superconductor--normal metal--$p_0$~superconductor ($sNp_0$) junction. The $p_0$-SC is a spinful $p$-wave superconductor with no topological channel: it has negative chemical potential ($\mu_p < 0$) and small Zeeman field $|V_Z| < |\mu_p|$. Its spectrum has a gap at $k = 0$ with value $|\mu_p| \pm |V_Z|$ where the smallest gap is~\cite{Kitaev,Read,Setiawan15} $\Delta_{p_0} = |\mu_p|- |V_Z|$ as shown in Fig.~\ref{Fig2}. In general, the current and conductance for SNS junctions involving $p_0$-SC, e.g., the $sNp_0$ junction discussed here, increase with the $p_0$-SC pairing potential $\Delta_{p}$.  Since the $p_0$-SC is essentially an insulator, the current for this junction is generally small and the SGS is strongly suppressed as can be seen in Fig.~\ref{Fig3}. At high voltages ($|V| \gg \Delta_s,\Delta_{p_0}$), the conductance approaches the conductance $G_N$ of the corresponding NN junction (which we define as the junction transparency throughout this paper). The current and conductance decrease with decreasing junction transparency $G_N$ as can be seen in Fig.~\ref{Fig3}. In the weak tunneling or small transparency limit where MAR is suppressed, the current starts to flow only when the voltage is $e|V| =  \Delta_s + \Delta_{p_0}$, i.e., the voltage where the superconducting gap edges of both \ssc and $p_0$-SC line up.  

\subsection{$sNp_1$ junction} 

\begin{figure}[t]
\capstart
\begin{center}\label{Fig4}
\includegraphics[width=\linewidth]{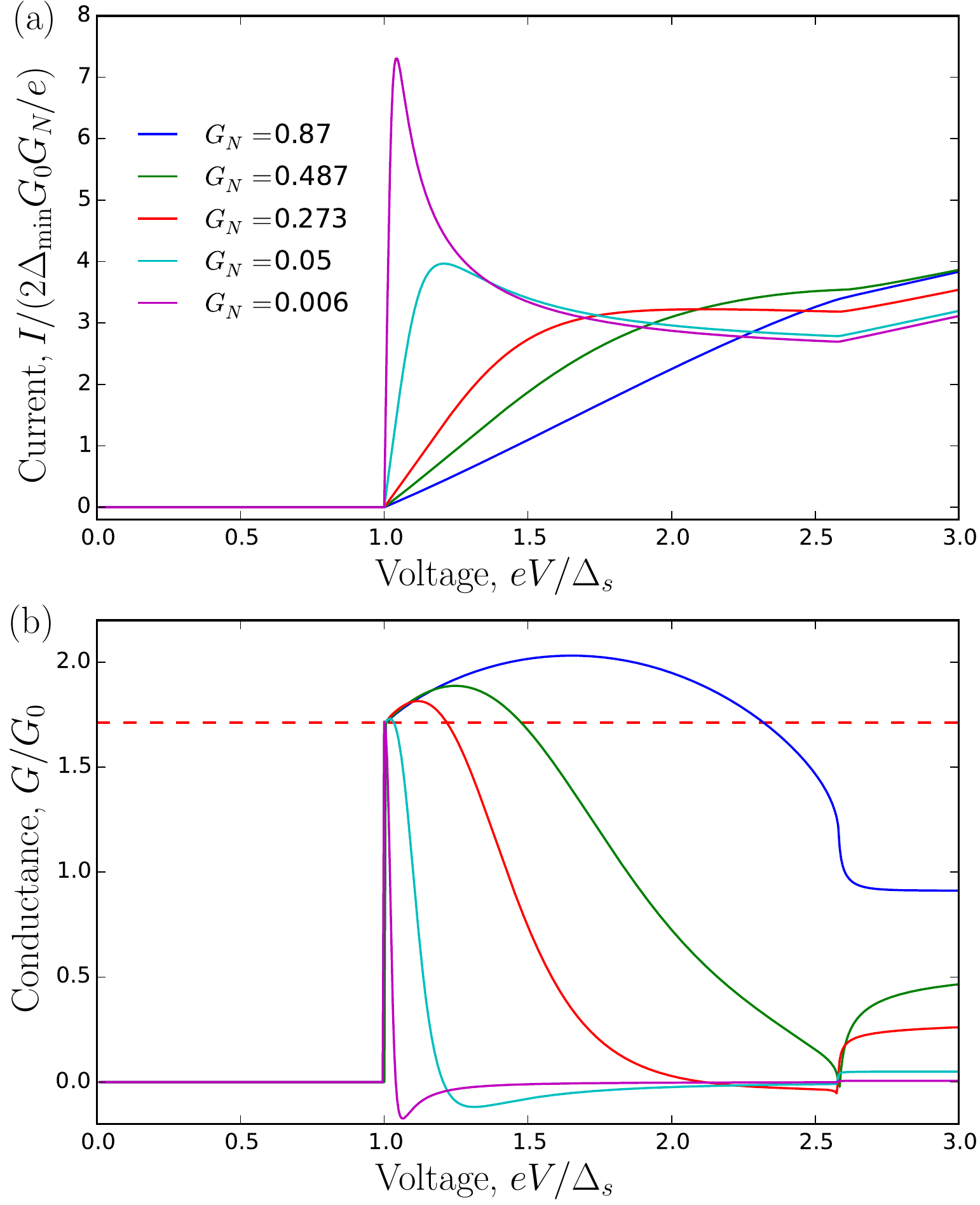}
\end{center}
\caption{(Color online) Plots of (a) dc current $I$ and (b) normalized differential conductance $G/G_0$ versus bias voltage $V$ for an $sNp_1$ junction with various values of transparencies $G_N$ in the limit of large Zeeman field ($V_Z = 2\mu_p$). The red dashed line at $G_M= (4-\pi)2e^2/h$ is the conductance value due to a single Andreev reflection from the MZM. The parameters used for the $s$SC are $\mu_{s} = 200$ K and $\Delta_{s} = 2.5$ K. The parameters used for the $p_1$-SC are $\mu_{p} = 20$ K, $V_{Z}= 40$ K, $\Delta_{p} = 0.0785$ eV\AA,  where the smallest gap is $\Delta_{p_1} = 4$ K. The smallest gap in the junction is $\Delta_{\mathrm{min}}$ = 2.5 K.} 
\end{figure}

\begin{figure}[h]
\capstart
\begin{center}\label{Fig5}
\includegraphics[width=\linewidth]{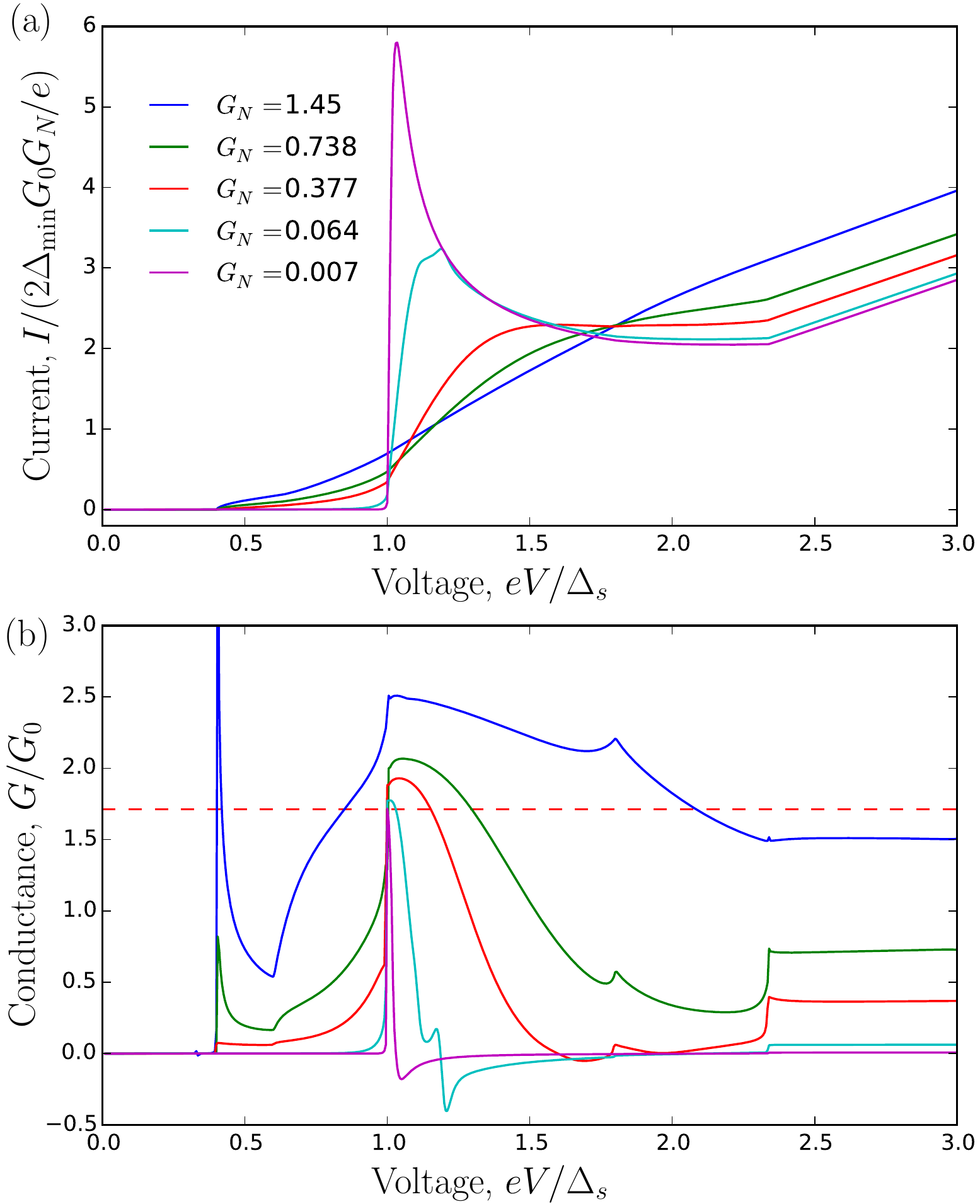}
\end{center}
\caption{(Color online) Plots of (a) dc current $I$ and (b) normalized differential conductance $G/G_0$ versus bias voltage $V$ for an $sNp_1$ junction with various values of transparencies $G_N$ in the limit of small Zeeman field ($V_Z = 1.1 \mu_p$). The red dashed line at $G_M= (4-\pi)2e^2/h$ is the conductance value due to a single Andreev reflection from the MZM. The parameters used for the $s$SC are $\mu_{s} = 200$ K and $\Delta_{s} = 2.5$ K. The parameters used for the $p_1$-SC are $\mu_{p} = 20$ K, $V_{Z}= 22$ K, $\Delta_{p} = 0.0785$ eV\AA,  where the gaps are 2 K and 3.4 K with the smallest gap for the $p_1$-SC being $\Delta_{p_1} = 2$ K. The smallest  gap in the junction is $\Delta_{\mathrm{min}}$ = 2 K.} 
\end{figure}

The $p_1$-SC has one topological channel with a pair of MZM: one at each end of a finite wire. The energy spectrum of the $p_1$-SC is given in Fig.~\ref{Fig2}(b). The plots of the current and conductance for the $sNp_1$ junction (i.e., with one isolated MZM in the junction) in the limit of large and small Zeeman field are plotted against bias voltage in Figs.~\ref{Fig4} and \ref{Fig5}, respectively. The conductance plots for the $sNp_1$ junction have already been given in Ref.~\cite{seti2016}; we include them here for completeness and, more importantly, a comparison with other SNS junctions. In the large Zeeman limit $[(|V_Z| -\mu_p) \gtrsim \mu_p]$, the $p_1$-SC is effectively a \emph{spinless} topological \psc~\cite{Kitaev,Read}. In this limit, MAR are totally suppressed and only single Andreev reflections are allowed for the $sNp_1$ junction because the \ssc allows only spin-singlet Andreev reflections, while the spinless \psc allows only spin-triplet Andreev reflections. This results in a step jump in the conductance from zero to the quantized value $G_M = (4-\pi)2e^2/h$ at the threshold voltage $e|V| = \Delta_s$~\cite{Peng15, Yeyati, seti2016} as shown in Fig.~\ref{Fig3}(b). The quantized value $G_M$  corresponds to the conductance due to a single Andreev reflection from the MZM which happens at the voltage when the BCS singularity and MZM are aligned. In this large Zeeman limit, since MAR are suppressed, the quantized value $G_M$ is robust against the junction transparency. The conductance, in general, decreases with decreasing junction transparency and for sufficiently small transparency, the conductance can become negative for voltages near the threshold voltage $e|V| = \Delta_s$. Our results for the $sNp_1$ junction in the large Zeeman limit, calculated using the scattering matrix formalism, are similar to those of the $s$-wave superconductor--normal metal--spinless $p$-wave superconductor junctions calculated using the Green's function formalism~\cite{Yeyati}. Recently, the conductance of the spinless $p$-wave superconductor has been measured using an $s$-wave superconducting tip in a scanning tunneling microscopy experiment~\cite{Ben}. The reported results are qualitatively consistent with our theoretical findings. In the limit of small Zeeman field $[(|V_Z| -\mu_p) \ll \mu_p]$, and when the junction transparency is not small, MAR are allowed. As a result, there is a finite current and conductance with SGS below the threshold voltage $e|V| = \Delta_s$. However, the current and conductance near zero voltage are zero due to the difference in the Andreev reflection spin-selectivity of the \ssc and MZM, i.e., the $s$SC allows spin-singlet Andreev reflection and the MZM favors spin-triplet Andreev reflection~\cite{jjhe,xinliu}. In this limit, due to MAR, the conductance at the voltage $e|V| = \Delta_s$ is no longer robust against increasing junction transparency. The current and conductance generally decrease with decreasing junction transparency. For sufficiently small transparency that only single Andreev reflection contributes, we recover $G(e|V| = \Delta_s)= G_M$.

\subsection{$sNp_2$ junction} 

\begin{figure}[h]
\capstart
\begin{center}\label{Fig6}
\includegraphics[width=\linewidth]{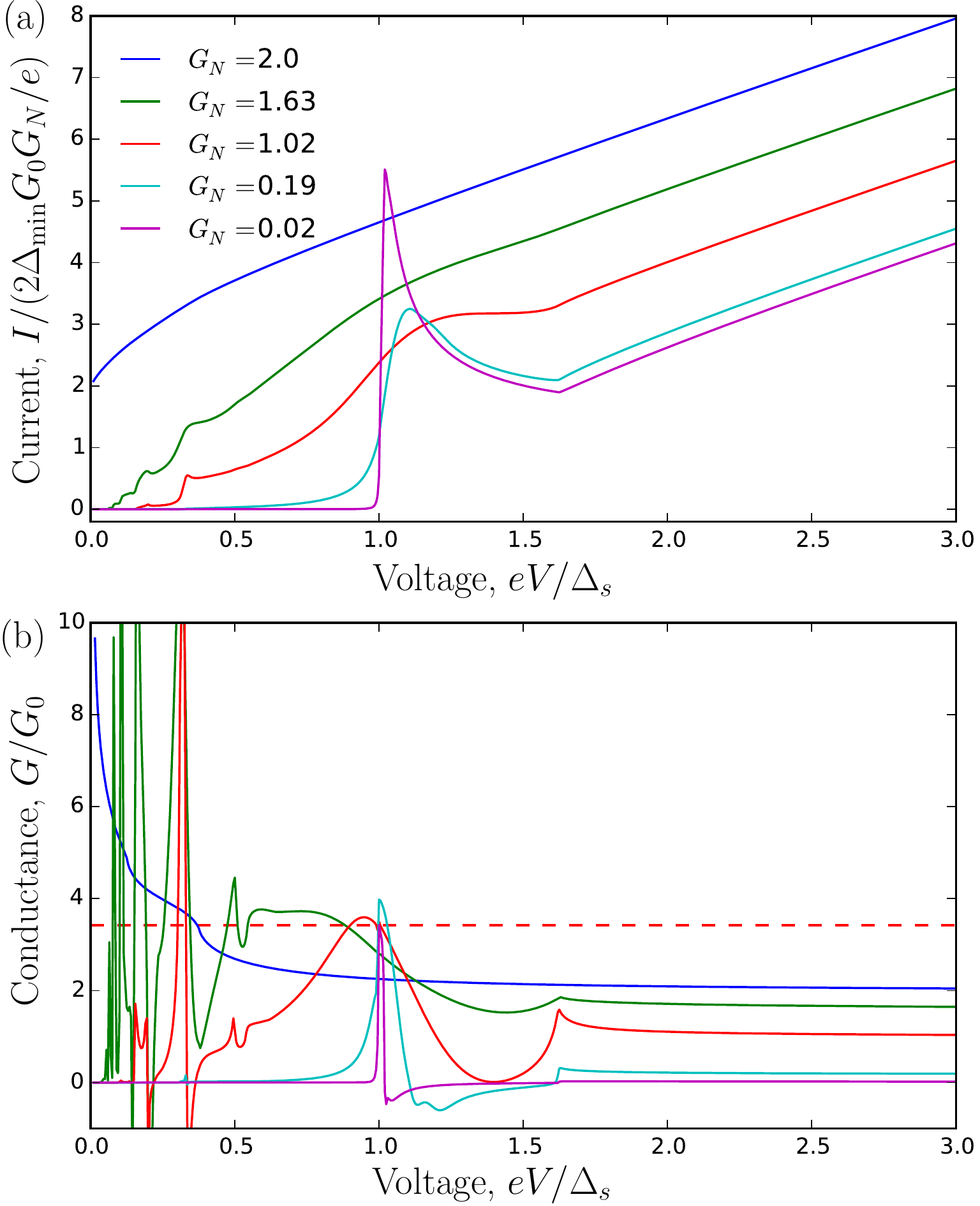}
\end{center}
\caption{(Color online) Plots of (a) dc current $I$ and (b) normalized differential conductance $G/G_0$ versus bias voltage $V$ for an $sNp_2$ junction with various values of transparencies $G_N$. The red dashed line at $2G_M= (4-\pi)4e^2/h$ is the conductance value due to a single Andreev reflection from two MZMs. The parameters used for the $s$SC are $\mu_{s} = 20$ K and $\Delta_{s} = 0.01$ K. The parameters used for the $p_2$-SC are $\mu_{p} = 20$ K, $V_{Z}= 0$ K, $\Delta_{p} = 2\times 10^{-4}$ eV\AA,  where the gap is $\Delta_{p_2} = 6.3\times 10^{-3}$ K. The smallest gap in the junction is $\Delta_{\mathrm{min}} = \Delta_{p_2} = 6.3\times 10^{-3}$ K.} 
\end{figure}

The $p_2$-SC has two topological channels, and thus two MZMs at each end of a finite wire. The energy spectrum for the $p_2$-SC is shown in Fig.~\ref{Fig2}(c). The current and conductance plots for the $sNp_2$ junction are depicted in Fig.~\ref{Fig6}. In the tunneling limit, the conductance for the $sNp_2$ junction develops a step jump from 0 to $2G_M$ at the threshold voltage $e|V| = \Delta_s$ due to single Andreev reflections from a Majorana Kramers pair, with each spin channel contributing a conductance of $G_M$. For large or intermediate transparencies, due to MAR, the conductance at $e|V| = \Delta_s$ is no longer quantized at $2G_M$ and there is an SGS in the current and conductance profile. In contrast to the $sNp_1$ junction where the current and conductance are zero near zero voltage, when the transparency is not small the current and conductance for the $sNp_2$ junction is nonzero near zero voltage. This is because unlike the case of the $sNp_1$ junction where there is only one MZM which facilitates the spin-triplet Andreev reflection in one spin channel, there are two MZMs in $sNp_2$ junctions facilitating Andreev reflections in both spin channels. As a result, the MAR are not suppressed near zero voltage. The SGS associated with MAR develops at specific voltages as given in Table~\ref{table2}. Similar to the conventional $s$-wave superconductor--normal--$s$-wave superconductor junction~\cite{Averin,Bagwell,Hurd}, in the perfectly transparent limit ($G_N = 2$), the current at small voltages for the $sNp_2$ junction asymptotically approaches 
\begin{equation}
I (V \rightarrow 0) = \frac{4 e \,\Delta_{\mathrm{min}}}{h},
\end{equation}
which corresponds to the transfer of a charge of $2e$ across the junction where $\Delta_{\mathrm{min}} = \mathrm{min} (\Delta_s,\Delta_{p_2})$ is the smallest gap in the junction.

\subsection{$p_2Np_2$ junction} 

\begin{figure}[h]
\capstart
\begin{center}\label{Fig7}
\includegraphics[width=\linewidth]{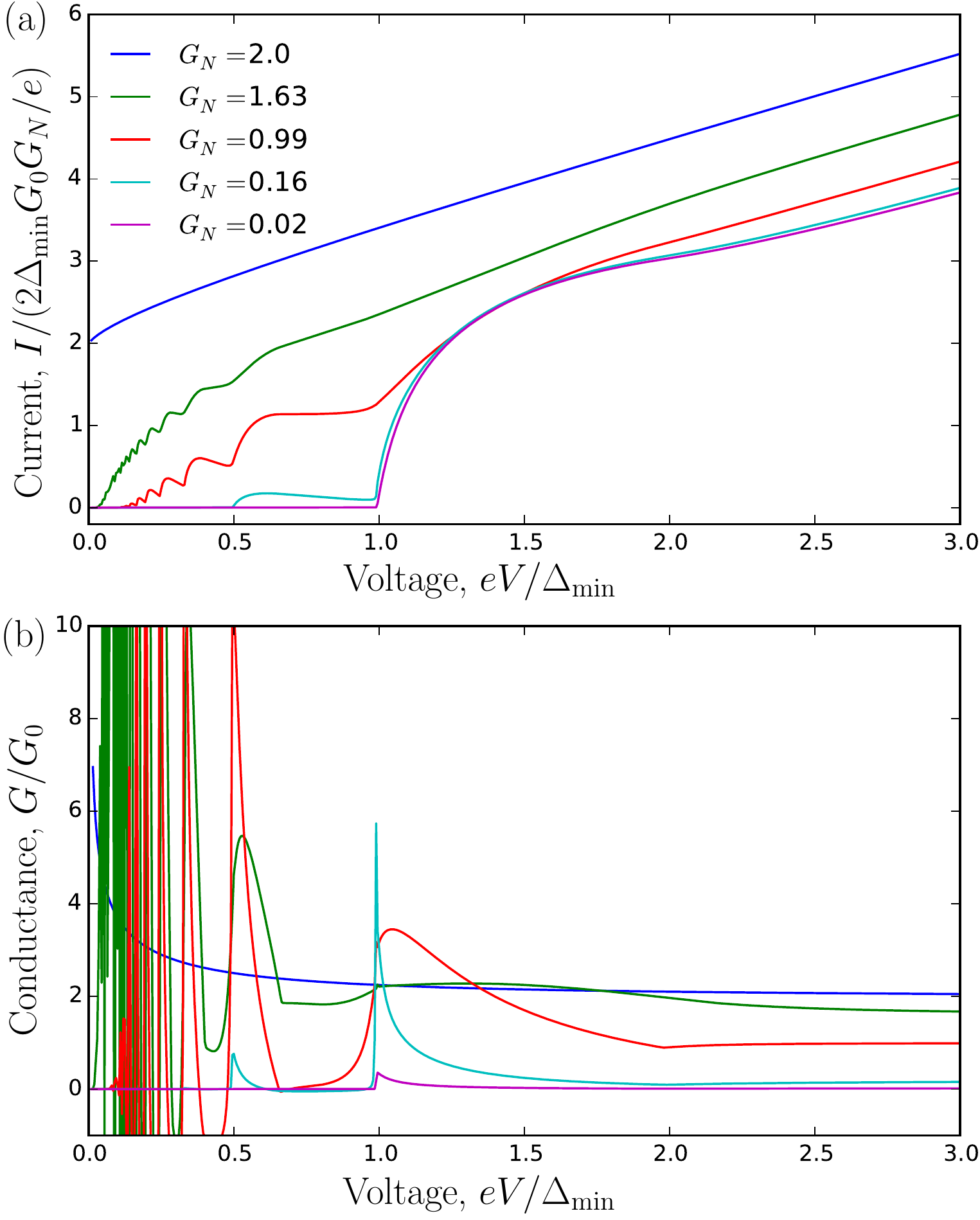}
\end{center}
\caption{(Color online) Plots of (a) dc current $I$ and (b) normalized differential conductance $G/G_0$ versus bias voltage $V$ for a $p_2Np_2$ junction with various values of transparencies $G_N$. The parameters used for both $p_2$-SCs are $\mu_p = 20$ K, $V_Z = 0$ K, $\Delta_p = 2\times 10^{-4}$ eV\AA, where  the gap is $\Delta_{p_2} = 6.3 \times 10^{-3}$ K. The smallest gap in the junction is $\Delta_{\mathrm{min}} = 6.3 \times 10^{-3}$ K.} 
\end{figure}
For a $p_2Np_2$ junction, both superconductors have two topological channels with two MZMs at each end (4 MZMs in the junction). The plots of the current and conductance for this junction are depicted in Fig.~\ref{Fig7}. In the perfectly transparent limit ($G_N = 2$), the current at small voltages asymptotically approaches $I (V \rightarrow 0) = 4 e \Delta_{\mathrm{min}}/h$, where $\Delta_{\mathrm{min}}$ is the smallest gap in the junction. This asymptotic value of the dc current is the same as the value obtained for the conventional $s$-wave-normal-$s$-wave superconductor junction~\cite{Averin,Bagwell,Hurd}. As $V\rightarrow 0$, the current in the $p_2Np_2$ junction is transferred via a Majorana Kramers pair where each of the MZMs transfers a charge of unit $e$ giving a total charge of $2e$, the same total amount of charge as that carried by a Cooper pair. As a result, the current $I (V \rightarrow 0)$ is the same as that for the conventional SNS junction~\cite{Averin,Bagwell,Zaikin}. Away from perfect transparency ($G_N \neq 2$), the dc current approaches zero as the voltage approaches zero.

The SGS associated with the MAR develops at specific voltages given in Table.~\ref{table2} where for the $p_2Np_2$ junction with symmetric gaps, the voltages are $|V| = \Delta_{p_2}/en$ as shown in Fig.~\ref{Fig7}.  The SGS is suppressed in the tunneling limit and the current becomes nonzero only when the voltage is above the threshold voltage $|V| = \Delta_{p_2}/e$, i.e., when the quasiparticles have sufficient energy to undergo single Andreev reflections from the MZMs. This is contrary to the case of the junction between two nontopological superconductors where the tunneling current can flow only when the voltage is above $|V| = 2\Delta/e$, i.e., when the gap edge of the unoccupied band lines up with that of the occupied band. Since the $p_2$-SC does not have a BCS singularity, the conductance at $|V| = \Delta_{p_2}/e$ in the tunneling limit is not quantized at $G_M$. Instead, it has a nonuniversal value, which decreases with decreasing junction transparency.


\subsection{$p_2Np_1$ junction} 

\begin{figure}[h]
\capstart
\begin{center}\label{Fig8}
\includegraphics[width=\linewidth]{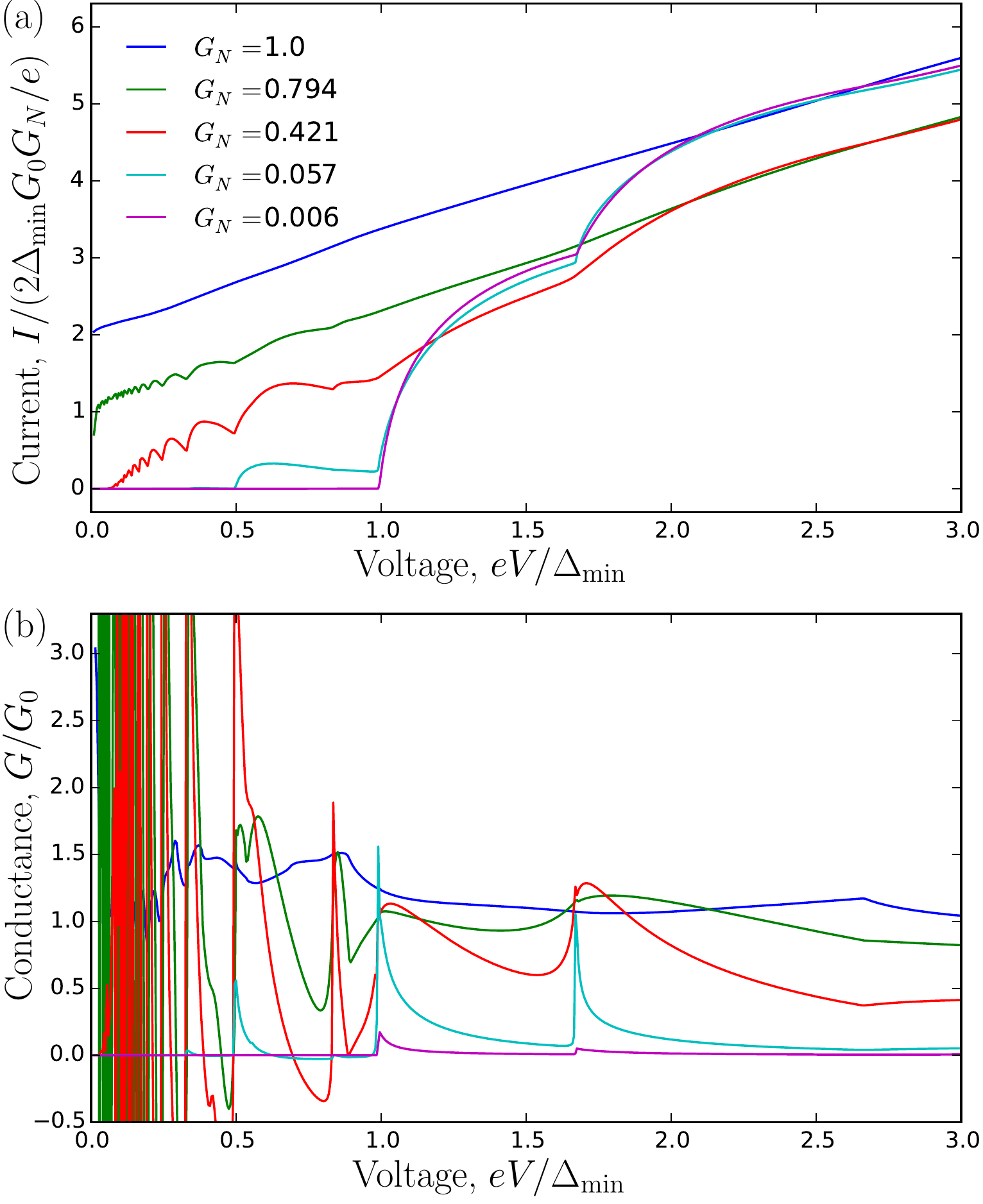}
\end{center}
\caption{(Color online) Plots of (a) dc current $I$ and (b) normalized differential conductance $G/G_0$ versus bias voltage $V$ for a $p_2Np_1$ junction with various values of transparencies $G_N$. The parameters used for the $p_1$-SC are $\mu_{p} = 20$ K, $\Delta_{p} = 2\times 10^{-4}$ eV\AA, and $V_{Z} = 40$ K, where the gap is $\Delta_{p_1}$ = 0.011 K. The parameters for the $p_2$-SC are $\mu_{p} = 20$ K, $\Delta_{p} = 2\times 10^{-4}$ eV\AA, $V_{Z} = 0$ K, where the gap is $\Delta_{p_2} = 6.3 \times 10^{-3}$ K. The smallest gap in the junction is $\Delta_{\mathrm{min}} = \Delta_{p_2} = 6.3 \times 10^{-3}$ K.} 
\end{figure}

The current and conductance for a $p_2Np_1$ junction are depicted in Fig.~\ref{Fig8}. For the $p_2Np_1$ junction in the perfectly transparent limit ($G_N = 1$), the current near zero voltage approaches $I (V \rightarrow 0) = 2 e \Delta_{\mathrm{min}}/h$,
which is half of the current for the $p_2Np_2$ or $s$-wave superconductor$-$normal metal$-s$-wave superconductor junction. The reason is that the $p_1$-SC has only one MZM which can transfer a charge in the unit of $e$ in one spin channel. The SGS appears at voltages given in Table~\ref{table2}. Since the $p_2Np_1$ junction considered here has asymmetric gap, the current and conductance in the weak tunneling limit develop jumps at the voltages $|V| = \Delta_{p_1}/e$ and $|V| = \Delta_{p_2}/e$ (which correspond to the voltages where the MZMs are aligned with the $p_1$-SC and $p_2$-SC superconducting gap edge). The conductance values at these jumps have nonuniversal values, which decrease with the junction transparency. 

\subsection{$p_1Np_1$ junction} 

\begin{figure}[h]
\capstart
\begin{center}\label{Fig9}
\includegraphics[width=\linewidth]{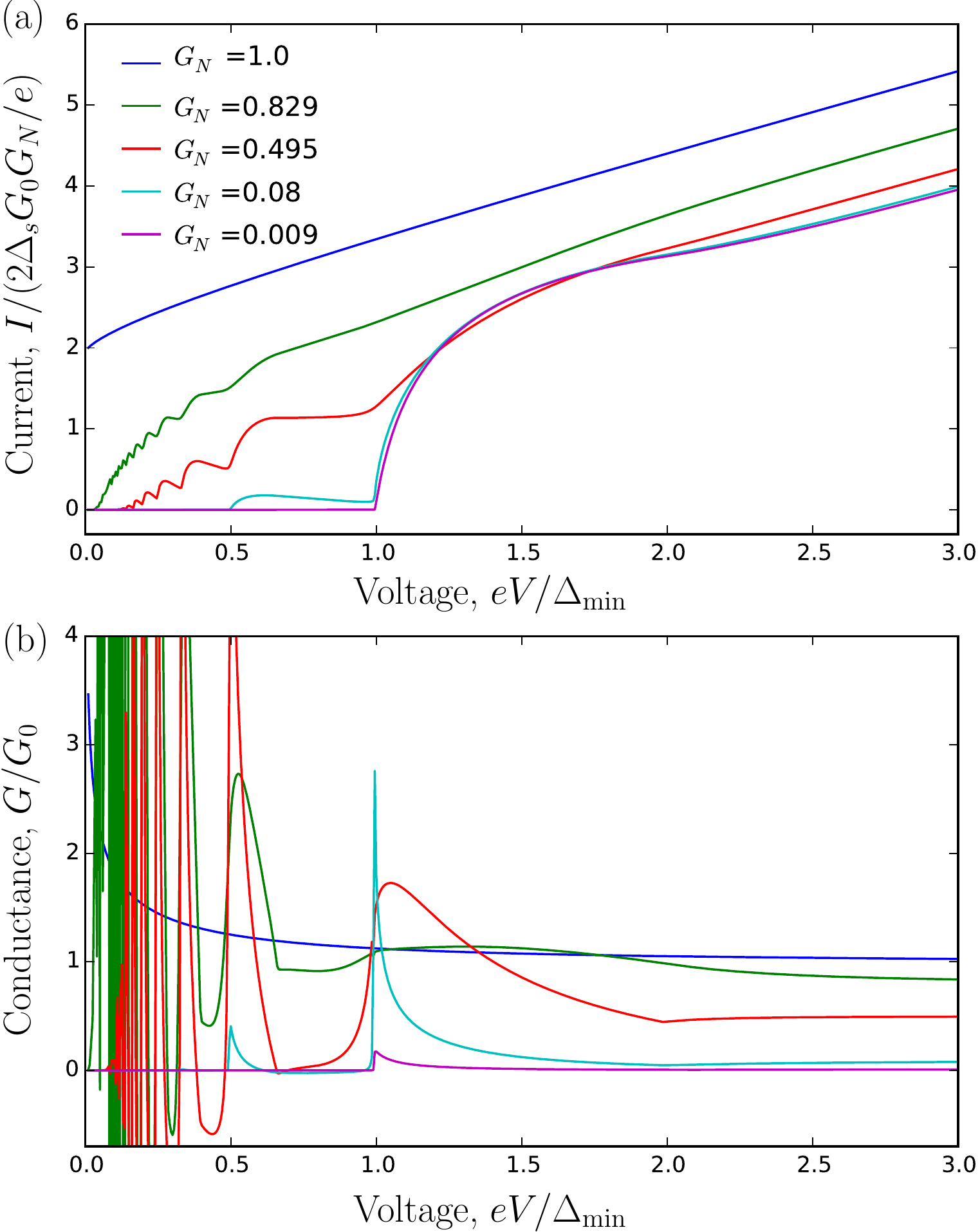}
\end{center}
\caption{(Color online) Plots of (a) dc current $I$ and (b) normalized differential conductance $G/G_0$ versus bias voltage $V$ for a $p_1Np_1$ junction with various values of transparencies $G_N$. The parameters used for both $p_1$-SCs are $\mu_{p} = 20$ K, $\Delta_{p} = 2\times 10^{-4}$ eV\AA, and $V_{Z} = 40$ K where the gap is $\Delta_{p_1}$ = 0.011 K.  The smallest gap in the junction is $\Delta_{\mathrm{min}} = \Delta_{p_1} = 0.011$ K.} 
\end{figure}
Figure~\ref{Fig9} displays the current and conductance plots for a $p_1Np_1$ junction. Similar to the $p_2Np_1$ junction, for a perfectly transparent $p_1Np_1$ junction ($G_N = 1$) the current at small voltages asymptotically approaches $I (V \rightarrow 0) = 2 e \Delta_{\mathrm{min}}/h$.
This is due to the fact that a charge of $e$ is transferred between the MZMs on both sides of the junction. For a symmetric $p_1Np_1$ as considered here, the SGS develops at voltages $|V| = \Delta_{p_1}/ne$. In the weak-tunneling limit, there is a step jump in the conductance at $|V| = \Delta_{p_1}/e$ to a nonuniversal value that decreases as the junction transparency decreases.

\subsection{$p_0Np_2$ junction}

\begin{figure}[h]
\capstart
\begin{center}\label{Fig10}
\includegraphics[width=\linewidth]{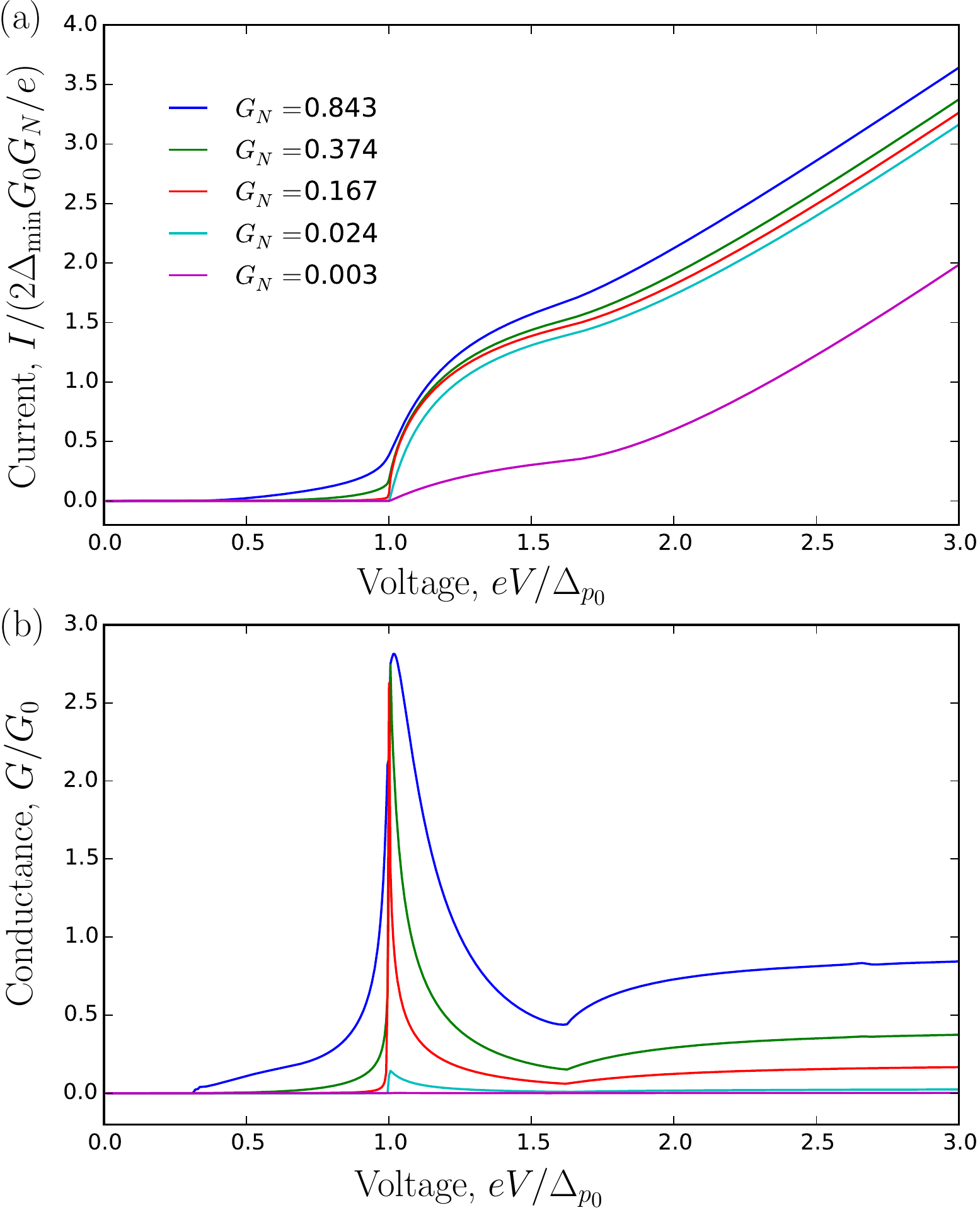}
\end{center}
\caption{(Color online) Plots of (a) dc current $I$ and (b) normalized differential conductance $G/G_0$ versus bias voltage $V$ for a $p_0Np_2$ junction with various values of transparencies $G_N$. The parameters used for the $p_0$-SC are  $\mu_{p} = -0.01$ K, $V_{Z}= 0$ K, $\Delta_{p} = 0.2$ eV\AA,  where the gap is $\Delta_{p_0} = 0.01$ K. The parameters used for the $p_2$-SC are $\mu_{p} = 20$ K, $V_{Z} = 0$ K, $\Delta_{p} = 2 \times 10^{-4}$ eV\AA, where the gap is $\Delta_{p_2} = 6.3 \times 10^{-3}$ K. The smallest gap in the junction is $\Delta_{\mathrm{min}} = 6.3 \times 10^{-3}$ K.} 
\end{figure}

The current and conductance plots for the $p_0Np_2$ junction are given in Fig.~\ref{Fig10}. The MAR for this junction are suppressed since $p_0$ is essentially an insulator. There is a conductance peak at $|V| = \Delta_{p_0}/e$, which corresponds to a single Andreev reflection from the MZMs. However, unlike the case of the $sNp_2$ junction, the tunneling conductance at the threshold voltage $|V| = \Delta_{p_0}/e$ assumes a nonquantized value which decreases with decreasing junction transparency. We note again that the MZM tunneling conductance quantization $G_M = (4-\pi)2e^2/h$ holds only if the superconducting probe has a BCS singularity (as derived in Ref.~\cite{Peng15}).

\subsection{$p_0Np_1$ junction} 
\begin{figure}[h]
\capstart
\begin{center}\label{Fig11}
\includegraphics[width=\linewidth]{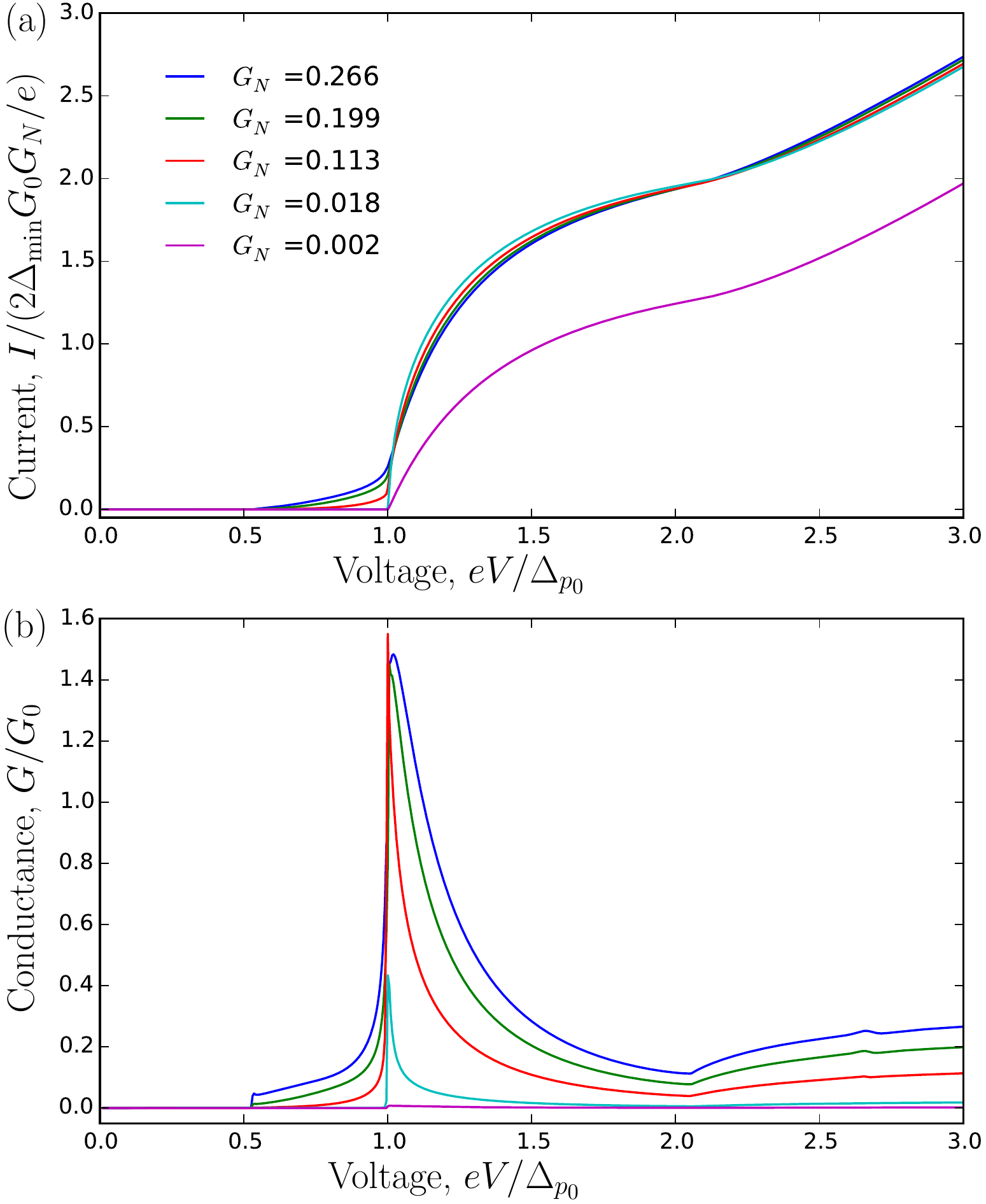}
\end{center}
\caption{(Color online) Plots of (a) dc current $I$ and (b) normalized differential conductance $G/G_0$ versus bias voltage $V$ for a $p_0Np_1$ junction with various values of transparencies $G_N$. The parameters used for the $p_0$-SC are  $\mu_{p} = -0.01$ K, $V_{Z}= 0$ K, $\Delta_{p} = 0.2$ eV\AA,  where the gap is $\Delta_{p_0} = 0.01$ K. The parameters used for the $p_1$-SC are $\mu_{p} = 20$ K, $V_{Z} = 40$ K, $\Delta_{p} = 2 \times 10^{-4}$ eV\AA, where the gap is $\Delta_{p_1} = 0.011$ K. The smallest gap in the junction is $\Delta_{\mathrm{min}} = 0.01$ K.} 
\end{figure}

The current and conductance plots for the $p_0Np_1$ junction are given in Fig.~\ref{Fig11}. The conductance for this junction looks similar to those of the $p_0Np_2$ junction. The MAR for this junction are suppressed and in the tunneling limit, the conductance has a step jump at the threshold voltage $e|V| = \Delta_{p_0}$ to a nonquantized value, which decreases with decreasing junction transparency.

\subsection{$p_0Np_0$ junction} 
\begin{figure}[h]
\capstart
\begin{center}\label{Fig12}
\includegraphics[width=\linewidth]{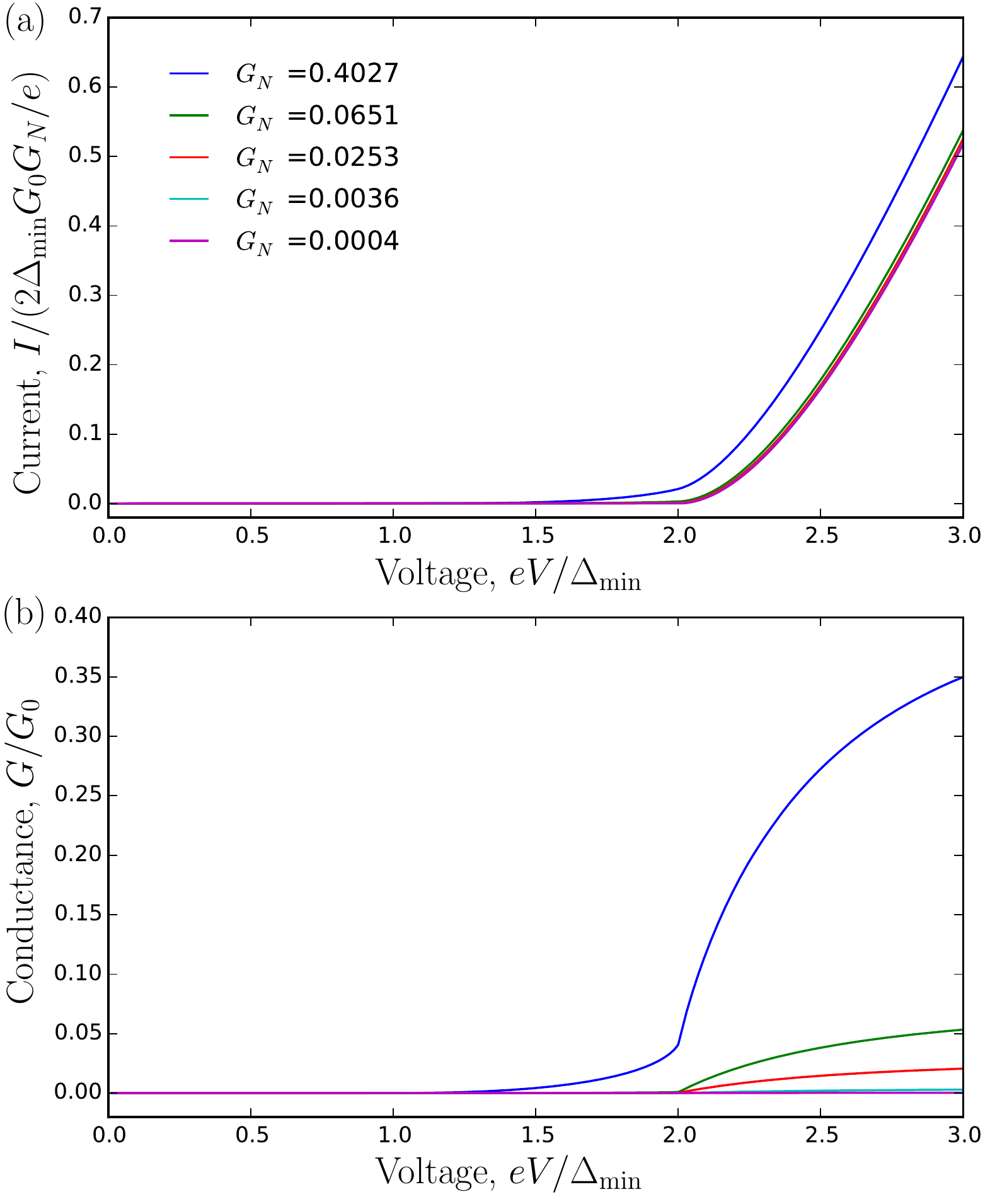}
\end{center}
\caption{(Color online) Plots of (a) dc current $I$ and (b) normalized differential conductance $G/G_0$ versus bias voltage $V$ for a $p_0Np_0$ junction with various values of transparencies $G_N$.  The parameters used for both $p_0$-SCs are  $\mu_p = -0.01$ K, $V_Z = 0$ K, $\Delta_p = 0.2$ eV\AA, where the gap is $\Delta_{p_0} = 0.01$ K. The smallest gap in the junction is $\Delta_{\mathrm{min}} = 0.01$ K.} 
\end{figure}

For the $p_0Np_0$ junction, the plots of the current and conductance versus the bias voltage are displayed in Fig.~\ref{Fig12}. Since the $p_0Np_0$ junction is essentially a junction between two insulators, the current and conductance for this junction are generally small and MAR are strongly suppressed. In the limit of small transparencies, the current and conductance for a symmetric $p_0Np_0$ junction rise to a nonzero value at $e|V| = 2\Delta_{p_0}$, i.e., when the density-of-state singularity of the occupied band of one $p_0$-SC is aligned with the singularity of the empty band of the other $p_0$-SC.

\section{Spin-orbit-coupled Superconducting Wire Junctions}\label{sec4}

\begin{figure}[h]
\capstart
\begin{center}\label{Fig13}
\includegraphics[width=\linewidth]{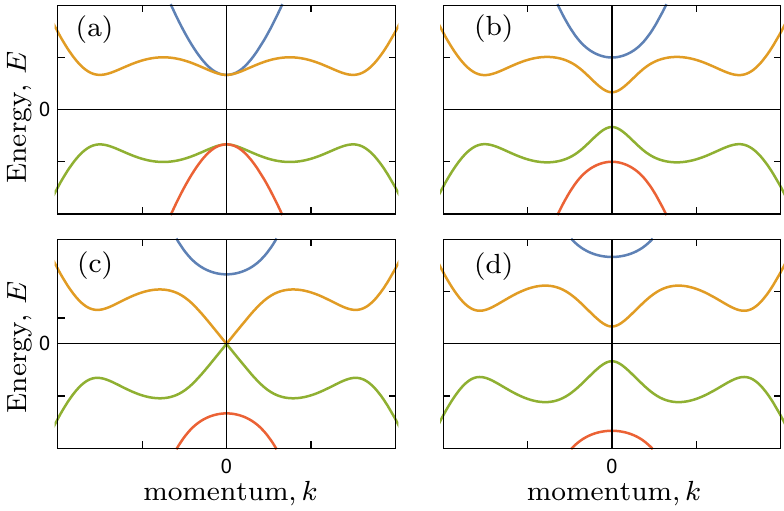}
\end{center}
\caption{(Color online) Energy spectrum of SOCSW for different parameter regimes: (a) $V_Z = 0$ (nontopological), (b) $V_Z  < \sqrt{\mu_0^2 + \Delta_0^2}$ (nontopological), (c) $V_Z = \sqrt{\mu_0^2 + \Delta_0^2}$ (transition), (d) $V_Z > \sqrt{\mu_0^2 + \Delta_0^2}$ (topological).} 
\end{figure}

Pure spinless or spinful $p$-wave topological superconductors as considered above do not exist in nature although they could be approximate models for some real systems.  It is, however, known that effective 1D or 2D topological superconductors closely mimicking spinless topological superconductors can be artificially engineered by combining $s$-wave superconductivity with spin-orbit coupling and Zeeman splitting~\cite{roman,oreg,jay09,jay10}. We therefore now consider a more physically realistic model for topological superconductors, namely, a spin-orbit-coupled 1D semiconducting nanowire placed in proximity to an $s$-wave superconductor in the presence of magnetic field~\cite{roman, oreg, jay10,alicea}. The $s$-wave superconductor induces superconductivity in the nanowire through proximity effect, and this proximity-induced nanowire superconductivity is converted to topological superconductivity by the Zeeman splitting in the nanowire (provided it is large enough to overcome the induced $s$-wave superconductivity) in the presence of spin-orbit coupling. The BdG Hamiltonian for the SOCSW is
\begin{align}
\mathcal{H}_{\mathrm {SOCSW}} = &\left(-\frac{\hbar^2\partial_x^2}{2m}-\mu_0\right)\tau_z \nonumber\\
&-i \alpha \partial_x\tau_z\sigma_y + V_Z\sigma_x + \Delta_0\tau_x, \label{eq:HamSOCSW}
\end{align}
where  $\mu_0$ is the chemical potential of the nanowire, $\alpha$ is the strength of the spin-orbit coupling, $V_Z$ is the Zeeman field, and $\Delta_0$ is the proximity-induced $s$-wave pairing potential. The Hamiltonian above is written in the same basis as that in Eq.~\eqref{eq:Hamilton}. The SOCSW can be tuned from the nontopological to the topological regime by simply changing the Zeeman field $V_Z$ or chemical potential $\mu_0$. The critical value $V_Z = \sqrt{\mu_0^2+\Delta_0^2}$ marks the topological quantum phase transition between the topologically trivial ($V_Z < \sqrt{\mu_0^2+\Delta_0^2}$) and topologically nontrivial phase ($V_Z > \sqrt{\mu_0^2+\Delta_0^2}$). In the topological regime, there is one MZM at each end of the nanowire (if the wire is long enough with well-separated MZMs, the system is in the topologically protected regime). The BdG spectrum of the SOCSW is given in Fig.~\ref{Fig13}. In what follows, we are going to denote the minimum gap in the SOCSW spectrum by $\Delta_{\mathrm{SOCSW}}$. Below we calculate the current and conductance of several SNS junctions between two SOCSWs where the SOCSW can be either in the nontopological or topological regime. The results given in the subsections below are our most relevant theoretical results for the currently ongoing MZM experiments in the literature, which mostly involve semiconductor nanowires.

\subsection{Nontopological--nontopological SOCSW junction} 

\begin{figure}[h]
\capstart
\begin{center}\label{Fig14}
\includegraphics[width=\linewidth]{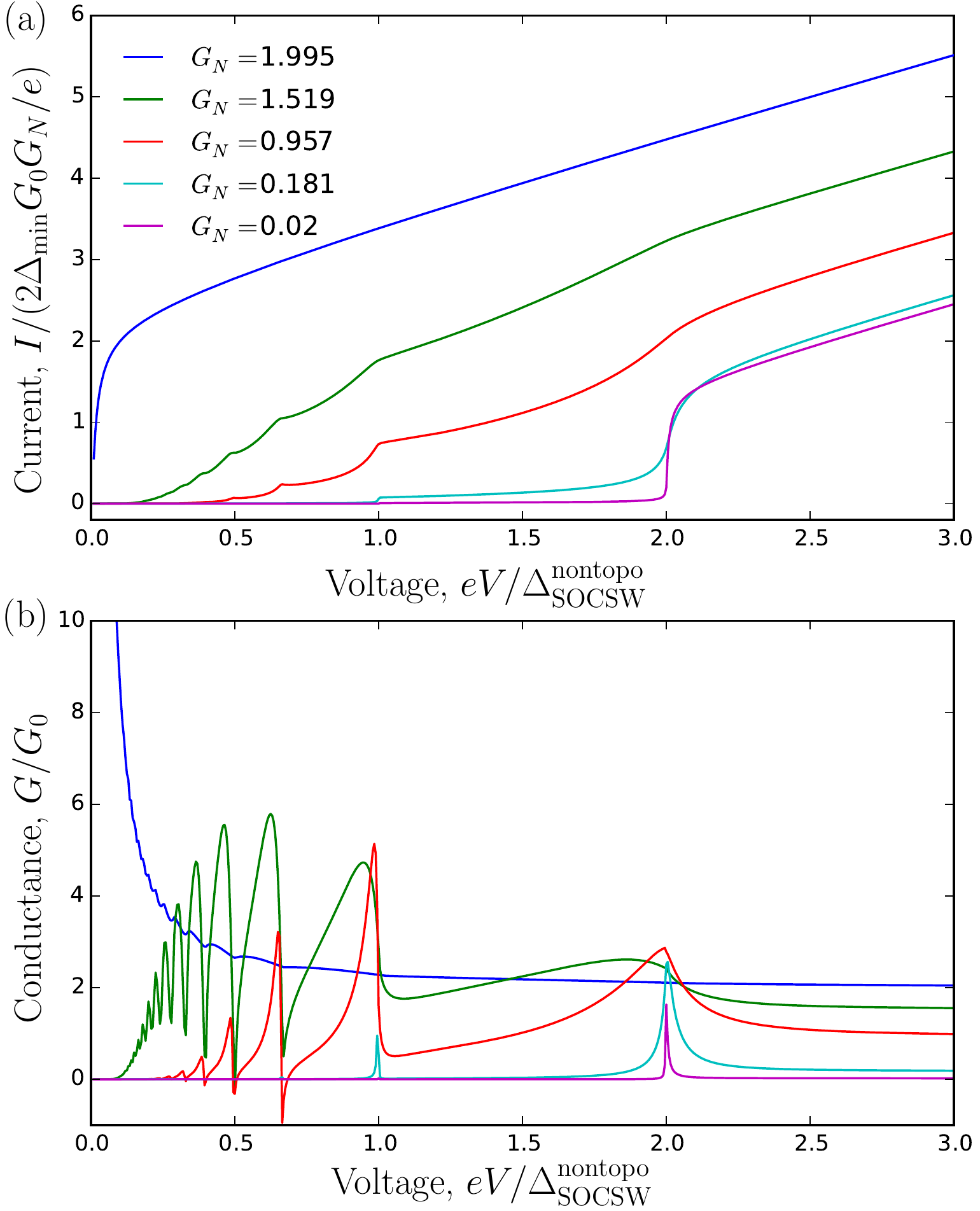}
\end{center}
\caption{(Color online) Plots of (a) dc current $I$ and (b) normalized differential conductance $G/G_0$ versus bias voltage $V$ for a nontopological--nontopological SOCSW junction with various values of transparencies $G_N$ and no Zeeman field. The parameters used for both SOCSWs are $\mu_0 =  0 $ K, $V_Z = 0$ K, $\Delta_0 = 0.01$ K, $\alpha = 0.5$ eV\AA, where the gap is $\dntsoc$ = 0.01 K. The smallest gap in the junction is $\Delta_{\mathrm{min}} = 0.01$ K.} 
\end{figure}

In this subsection, we consider the junction between two SOCSWs where both of them are in the nontopological regime (i.e., $V_Z < \sqrt{\mu_0^2 + \Delta_0^2}$). As shown in Fig.~\ref{Fig14}, the current and conductance of this junction with no Zeeman field ($V_Z = 0$) is the same as that of an $s$-wave superconductor--normal metal--$s$-wave superconductor junction~\cite{Averin,Hurd,Bagwell}. The SGS for the symmetric nontopological--nontopological SOCSW junction occurs at voltages $|V| = 2\dntsoc/ne$. For a perfectly transparent junction ($G_N = 2$), the current at small voltages approaches the value
\begin{equation}
I(V \rightarrow 0) = \frac{4e\Delta_{\mathrm{min}}}{h}.
\end{equation}
In the limit of small transparency the current and conductance develop a step jump at $|V| = 2\dntsoc/e$ for junctions with symmetric gaps.

\begin{figure}[h]
\capstart
\begin{center}\label{Fig15}
\includegraphics[width=\linewidth]{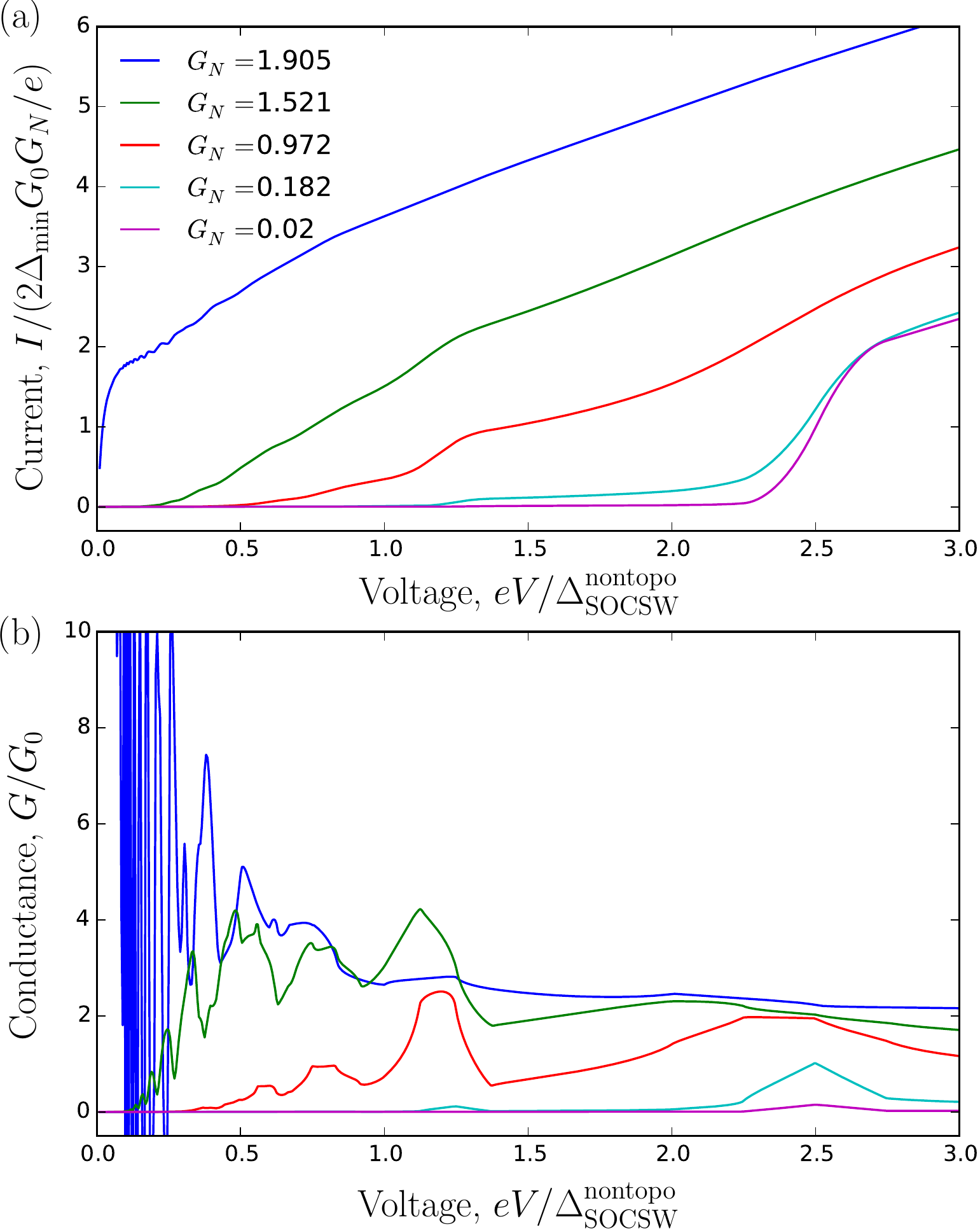}
\end{center}
\caption{(Color online) Plots of (a) dc current $I$ and (b) normalized differential conductance $G/G_0$ versus bias voltage $V$ for a nontopological--nontopological SOCSW junction with various values of transparencies $G_N$ and finite Zeeman field. The parameters used for both SOCSWs are $\mu_0 =  0 $ K, $V_Z = 0.002$ K, $\Delta_0 = 0.01$ K, $\alpha = 0.5$ eV\AA, where the gap is $\dntsoc$ = 0.008 K. The smallest gap in the junction is $\Delta_{\mathrm{min}} = 0.008$ K.} 
\end{figure}

Figure~\ref{Fig15} shows the current and conductance for the nontopological--nontopological SOCSW junction in the presence of Zeeman field. Increasing the Zeeman field smooths out the SGS. In the limit of small transparencies, the conductance has a smooth rise from zero instead of a step jump at the threshold voltage.

\subsection{Nontopological--topological SOCSW junction} 

\begin{figure}[h]
\capstart
\begin{center}\label{Fig16}
\includegraphics[width=\linewidth]{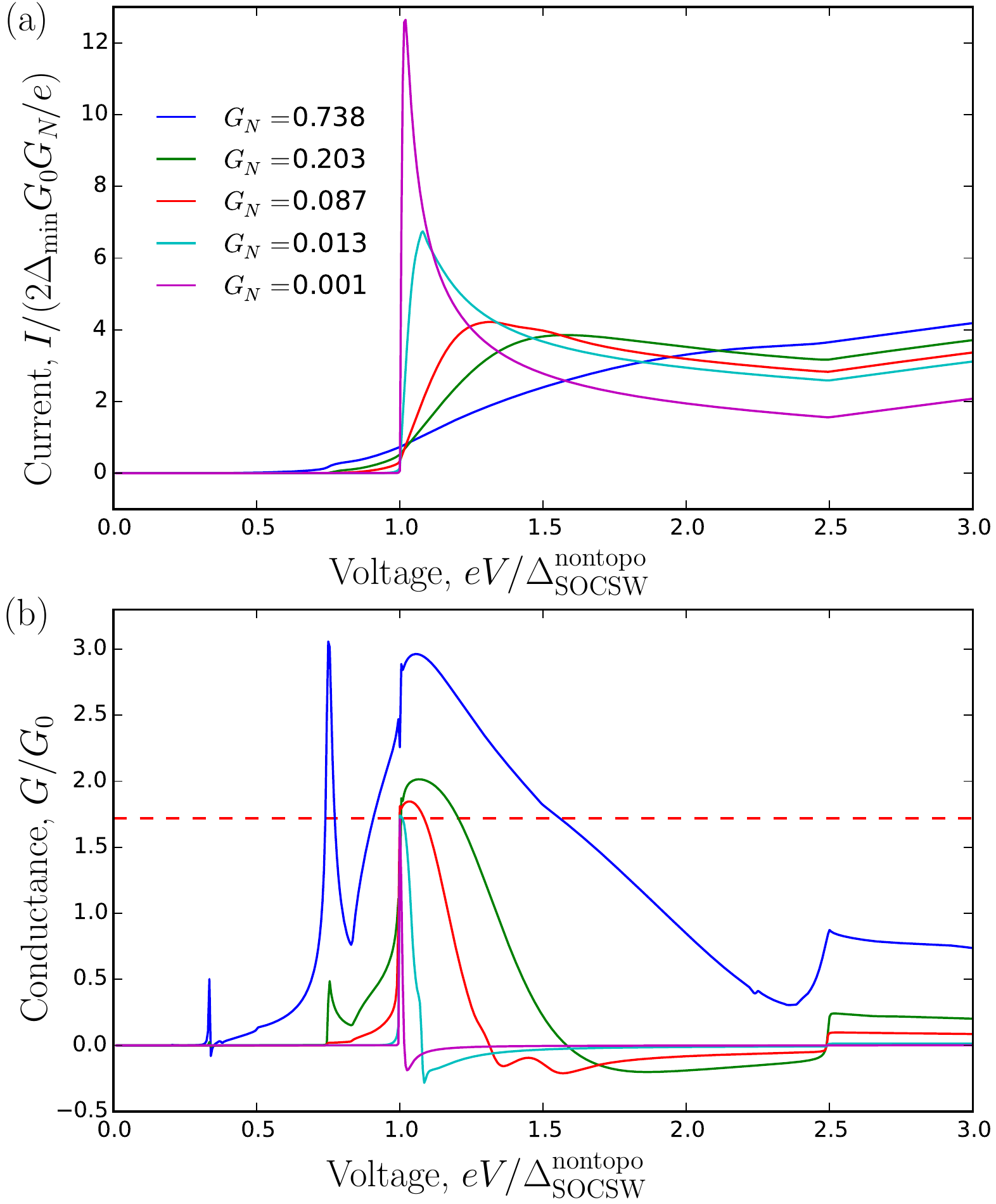}
\end{center}
\caption{(Color online) Plots of (a) dc current $I$ and (b) normalized differential conductance $G/G_0$ versus bias voltage $V$ for a nontopological--topological SOCSW junction with various values of transparencies $G_N$. The red dashed line at $G_M= (4-\pi)2e^2/h$ is the conductance value due to a single Andreev reflection from the MZM. The nontopological SOCSW is not subjected to any Zeeman field and the topological superconductor has a small Zeeman field. The parameters used for the nontopological SOCSW are $\mu_0 =  0 $ K, $V_Z = 0$ K, $\Delta_{0} = 0.5$ K, $\alpha = 0.5$ eV\AA, where $\dntsoc = 0.5$ K. The parameters used for the topological SOCSW are $\mu_0 =  0$ K, $V_Z = 15.0$ K, $\Delta_{0} = 10.0$ K, $\alpha = 0.05$ eV\AA, where the gap is $\dtsoc = 0.75$ K. The smallest gap in the junction is $\Delta_{\mathrm{min}} = 0.5$ K.} 
\end{figure}

\begin{figure}[h]
\capstart
\begin{center}\label{Fig17}
\includegraphics[width=\linewidth]{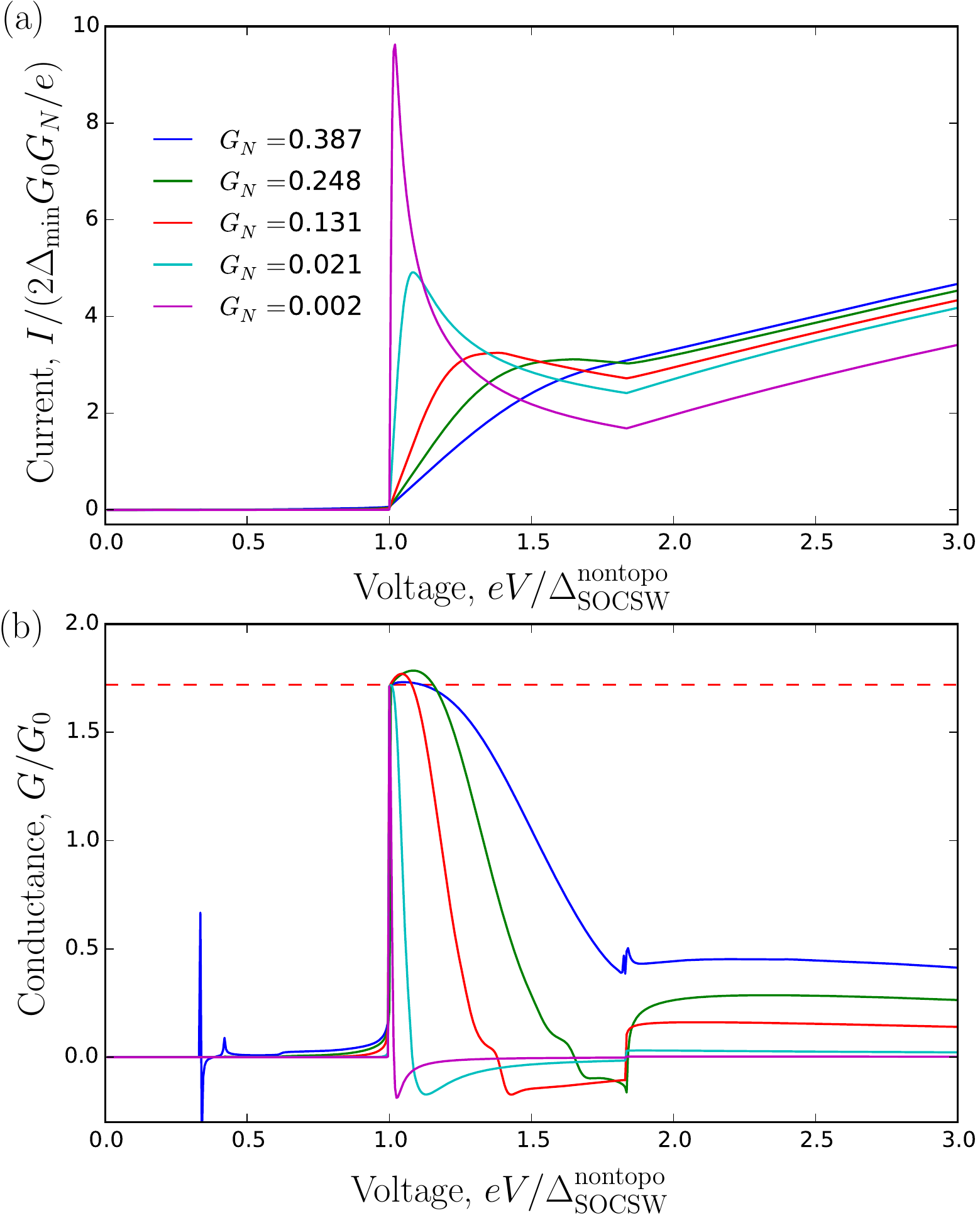}
\end{center}
\caption{(Color online) Plots of (a) dc current $I$ and (b) normalized differential conductance $G/G_0$ versus bias voltage $V$ for a nontopological--topological SOCSW junction with various values of transparencies $G_N$. The red dashed line at $G_M= (4-\pi)2e^2/h$ is the conductance value due to a single Andreev reflection from the MZM. The nontopological SOCSW is not subjected to any Zeeman field and the topological superconductor has a large Zeeman field. The parameters used for the nontopological SOCSW are $\mu_0 =  0 $ K, $V_Z = 0$ K, $\Delta_{0} = 0.5$ K, $\alpha = 0.5$ eV\AA, where $\dntsoc = 0.5$ K. The parameters used for the topological SOCSW are $\mu_0 =  0$ K, $V_Z = 60.0$ K, $\Delta_{0} = 10.0$ K, $\alpha = 0.05$ eV\AA, where the gap is $\dtsoc = 0.42$ K. The smallest gap in the junction is $\Delta_{\mathrm{min}} = 0.42$ K.} 
\end{figure}

\begin{figure}[h]
\capstart
\begin{center}\label{Fig18}
\includegraphics[width=\linewidth]{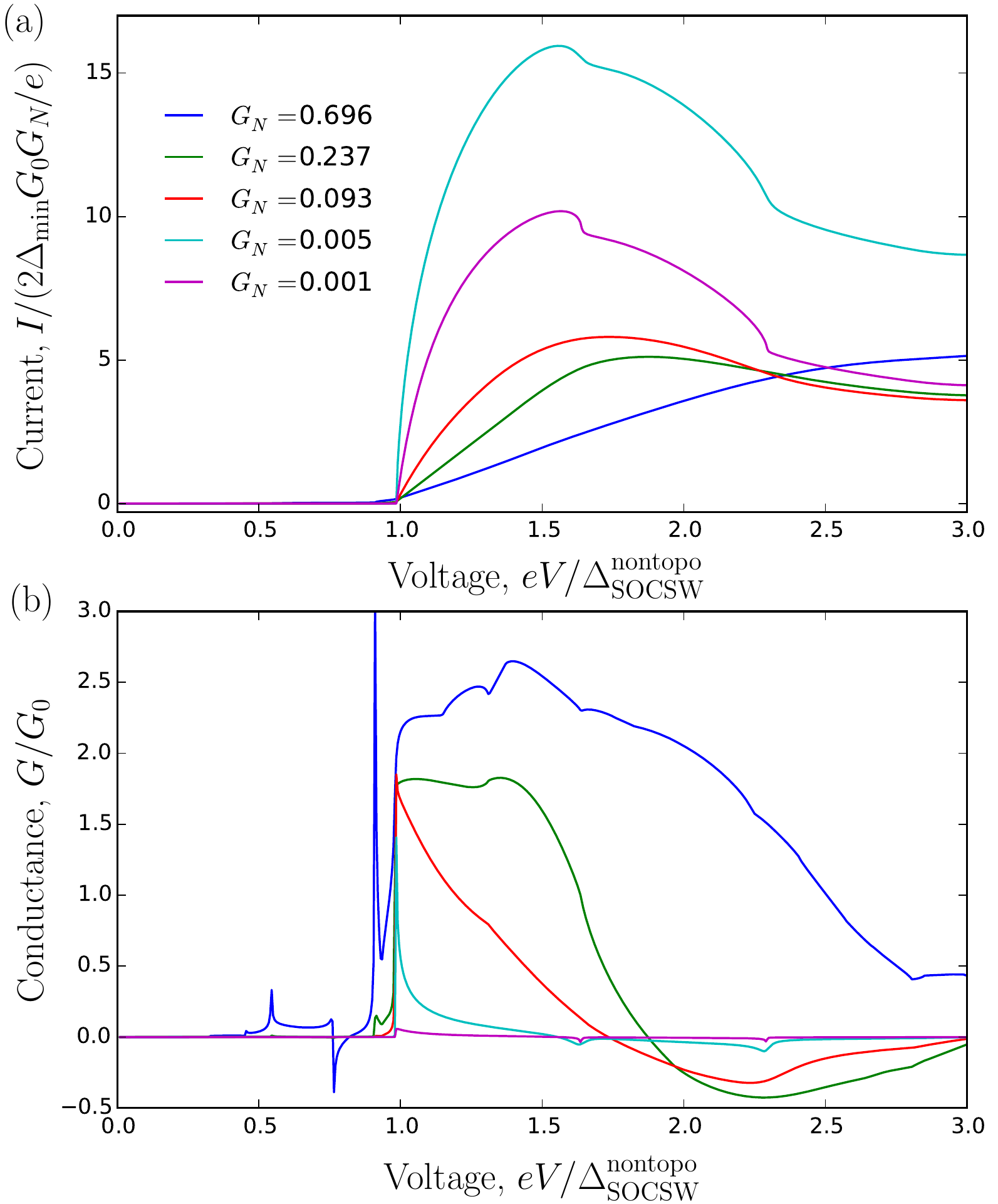}
\end{center}
\caption{(Color online) Plots of (a) dc current $I$ and (b) normalized differential conductance $G/G_0$ versus bias voltage $V$ for a nontopological--topological SOCSW junction with various values of transparencies $G_N$. The nontopological SOCSW  has a finite Zeeman field and the topological superconductor has a small Zeeman field. Note that the MZM tunneling conductance is not quantized at $G_M = (4-\pi)2e^2/h$. The parameters used for the nontopological SOCSW are $\mu_0 =  0 $ K, $V_Z = 0.2$ K, $\Delta_{0} = 0.5$ K, $\alpha = 0.5$ eV\AA, where $\dntsoc = 0.3$ K. The parameters used for the topological SOCSW are $\mu_0 =  0$ K, $V_Z = 15.0$ K, $\Delta_{0} = 10.0$ K, $\alpha = 0.05$ eV\AA, where the gap is $\dtsoc = 0.75$ K. The smallest gap in the junction is $\Delta_{\mathrm{min}} = 0.3$ K.} 
\end{figure}

Here, we consider junctions between a nontopological and a topological SOCSW. The current and conductance for such junctions are given in Figs.~\ref{Fig16}-\ref{Fig18}. We first consider the case of the junction with the nontopological SOCSW having no Zeeman field where the energy spectrum for this nontopological SOCSW has the minimum gap at the Fermi momentum with a BCS singularity [as shown in Fig.~\ref{Fig13}(a)]. As shown in Figs.~\ref{Fig16} and ~\ref{Fig17}, the conductance in the tunneling limit for this junction develops a step jump from 0 to $G_M = (4-\pi)2e^2/h$ at the gap-bias voltage $e|V| = \Delta_{\mathrm{SOCSW}}^{\mathrm{nontopo}}$ similar to the case of $sNp_1$ junction. This quantized value $G_M$ is due to a single Andreev reflection from the MZM of an electron coming from the gap edge with a BCS singularity. In the limit where the Zeeman field in the topological SOCSW is small, for intermediate and large transparencies, there are MAR and the conductance below the voltage $e|V| = \Delta_{\mathrm{SOCSW}}^{\mathrm{nontopo}}$ is nonzero except for small voltages (see Fig.~\ref{Fig16}). Near zero voltage, the current and conductance vanish due to a mismatch in the Andreev reflection spin-selectivity between the nontopological SOCSW and the MZM. In the limit of large Zeeman field in the topological SOCSW, where the MAR are suppressed and only single Andreev reflections are allowed, the conductance for this junction develops a step jump from 0 to $G_M = (4-\pi)2e^2/h$ independent of the junction transparency. We note that this result is similar to the case where the nontopological SOCSW is replaced by an $s$-wave superconductor~\cite{seti2016}.

For the case where there is Zeeman field in the nontopological superconductor, the gap edge of the superconductor no longer has the BCS singularity. As a result, the MZM tunneling conductance measured using this nontopological superconductor will \emph{not} be quantized at $G_M$ for the gap-bias voltage $e|V| = \Delta_{\mathrm{SOCSW}}^{\mathrm{nontopo}}$. Instead, the tunneling conductance assumes a nonuniversal value which decreases with decreasing junction transparency as shown in Fig.~\ref{Fig18}.

\subsection{Topological--topological SOCSW junction} 
\begin{figure}[h]
\capstart
\begin{center}\label{Fig19}
\includegraphics[width=\linewidth]{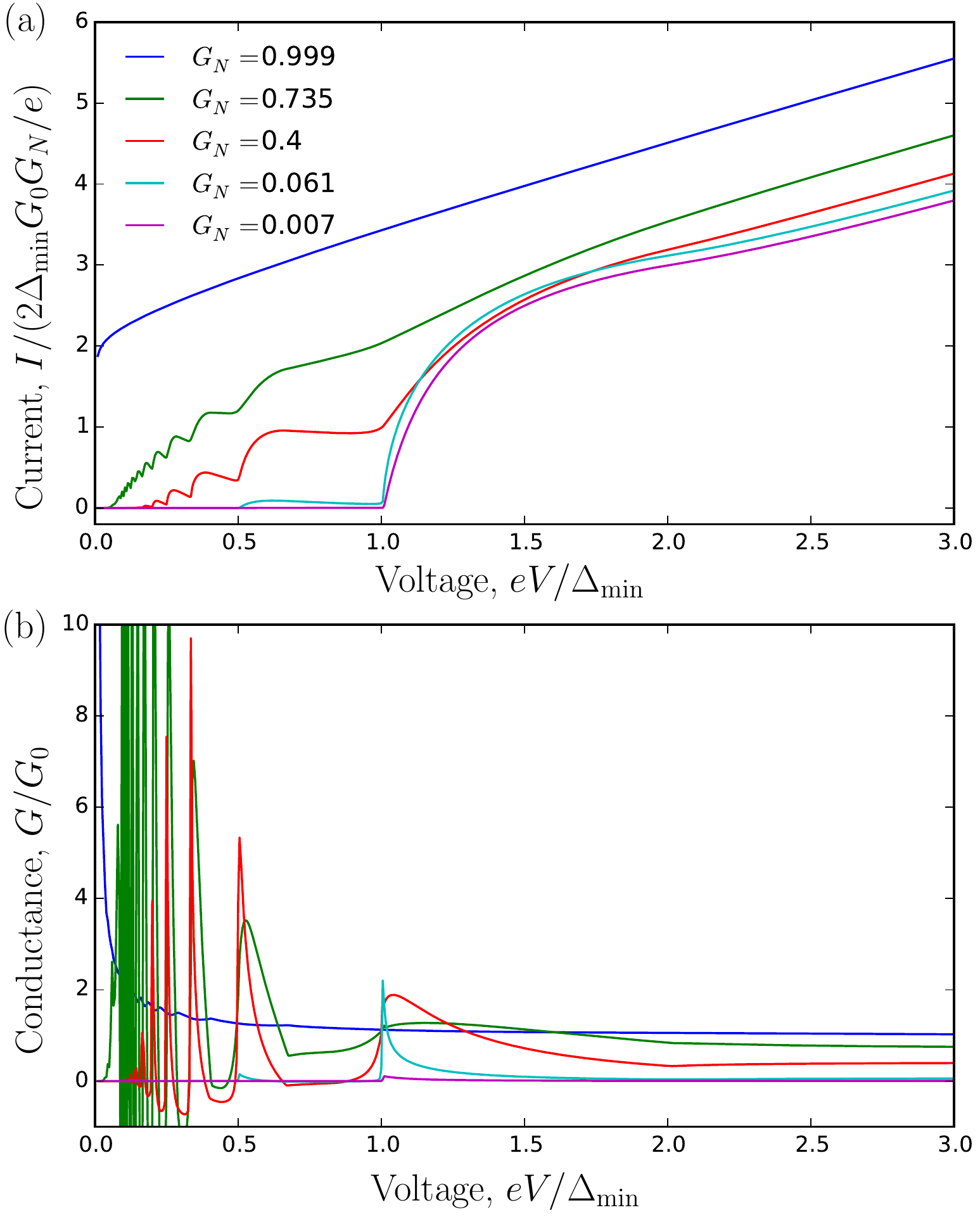}
\end{center}
\caption{(Color online) Plots of (a) dc current $I$ and (b) normalized differential conductance $G/G_0$ versus bias voltage $V$ for a topological--topological SOCSW junction with various values of transparencies $G_N$. The parameters used for both SOCSWs are $\mu_0 =  0$ K, $V_Z = 15$ K, $\Delta_0 = 1.17$ K, $\alpha = 0.05$ eV\AA, where the gap is $\dtsoc = 0.01$ K. The smallest gap in the junction is $\Delta_{\mathrm{min}} = 0.01$ K.} 
\end{figure}

The current and conductance plots for a topological--topological SOCSW junction are shown in Fig.~\ref{Fig19}. Our results for this junction, calculated using the scattering matrix formalism, are identical to previous results for the same SNS junction calculated using a Green's function method~\cite{aguado} and similar to the results obtained in Ref.~\cite{Badiane} for a topological Josephson junction between superconductors connected through the helical edge states of a 2D topological insulator in the presence of a magnetic barrier.

Similar to the $p_1Np_1$ junction, in the limit of perfect transparency ($G_N = 1$), the current for a topological--topological SOCSW junction asymptotically approaches 
\begin{equation}
I(V\rightarrow 0) = \frac{2e\Delta_{\mathrm{min}}}{h},
\end{equation}
which is half the value of the current in the conventional SNS junction. The reason is because there is only one MZM at both sides of the junction which transfer charges in unit of $e$. The SGS for this junction happens at voltages $|V| = \dmin/ne$. In the weak tunneling limit, there is a step jump in the conductance at $|V| = \dmin/e$. We note, however, that since there is no BCS singularity in the superconducting lead, the conductance at the voltage $|V| = \dmin/e$ is not quantized at $G_M = (4-\pi)2e^2/h$.

\section{Andreev Bound States}\label{sec5}

\begin{figure*}
\capstart
\begin{center}\label{Fig20}
\includegraphics[width=\linewidth]{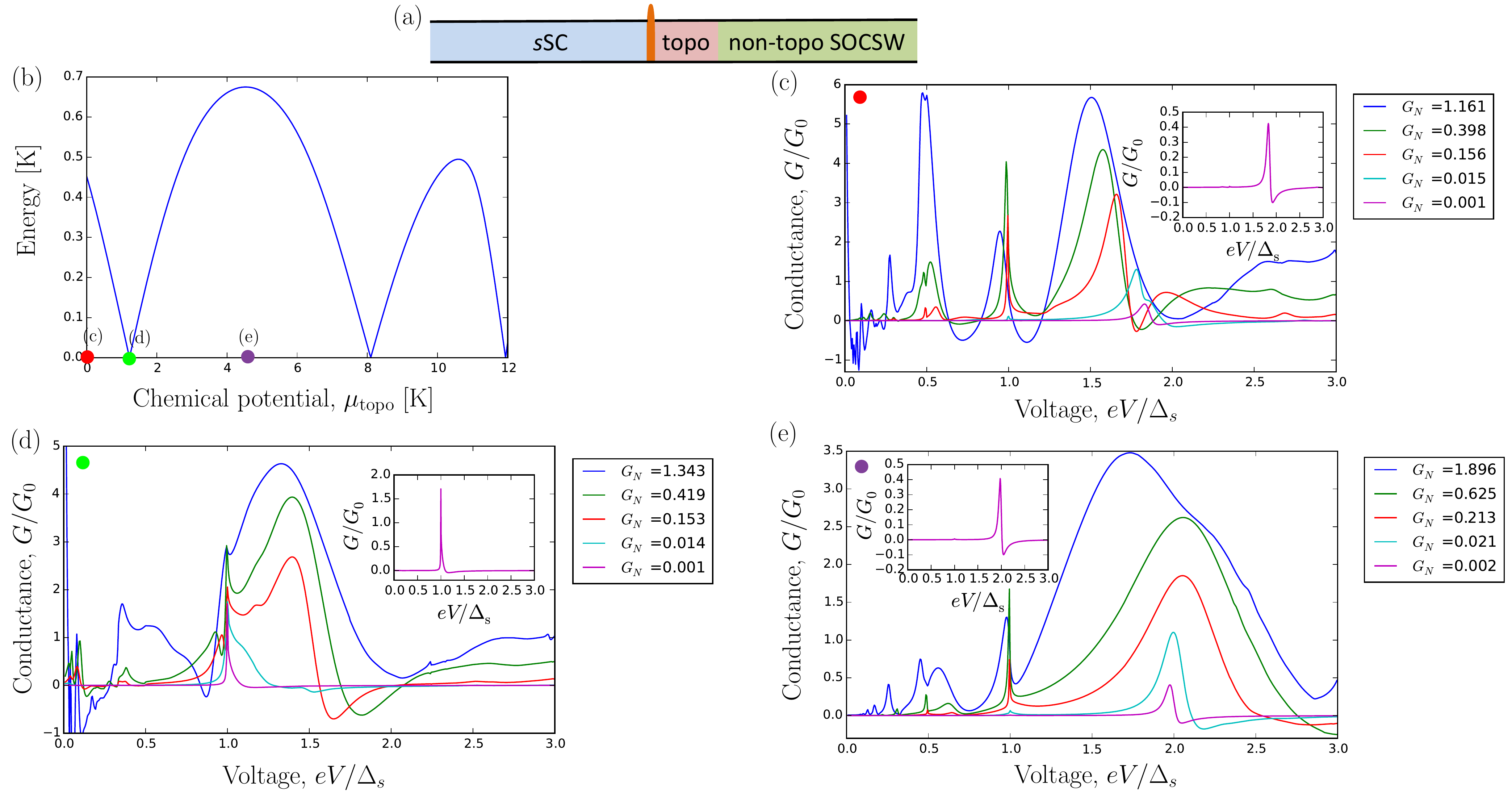}
\end{center}
\caption{(Color online) (a) Schematic diagram of an \ssc--SOCSW junction with a pair of ABS (one at each end of the topological region). The chemical potential of the topological and nontopological regions are $|\mu_{\mathrm{topo}}| < \sqrt{V_Z^2 - \Delta_0^2}$ and $|\mu_{\mathrm{nontopo}}| > \sqrt{V_Z^2 - \Delta_0^2}$, respectively. The parameters used for the \ssc are $\mu_s = 50$ K, and $\Delta_s =$ 0.67 K. The SOCSW parameters are $\mu_{\mathrm{nontopo}} =$ 211.18 K, $V_Z =$ 15 K, $\Delta_0 =$ 10 K, $\alpha =$ 0.05 eV\AA, and length of the topological region, $L_{\mathrm{topo}} = 0.6$ $\mu$m. We use a dissipation term $i\Gamma\tau_0 \otimes \sigma_0$ in the BdG Hamiltonian of both the left and right superconductors with a dissipation strength $\Gamma = 0.05 K$ to broaden the van-Hove singularity. (b) The energy of the Andreev bound state closest to zero energy versus the chemical potential $\mu_{\mathrm{topo}}$ in the topological region. The red, green and purple dots indicate the value of the topological chemical potential used in (c),(d), and (e), respectively. Normalized differential conductance $G/G_0$ for the SOCSW for several chemical potential values in the topological region: (c) $\mu_{\mathrm{topo}} =$ 0 K, (d) $\mu_{\mathrm{topo}} =$ 1.697 K, and (e) $\mu_{\mathrm{topo}} =$ 4.5 K. Inset: the ABS conductance in the weak tunneling limit which is the conductance for the smallest transparency in the main plot.} 
\end{figure*}

In this section we compare the conductance of an MZM with that of an ABS. We mention that the possible existence of ABS in the system can never be ruled out \textit{a priori}, and it is therefore important to take into account their possible effects on transport properties. In particular, we consider the ABS that may arise in the SOCSW model with a finite topological region and a semi-infinite nontopological region as shown in the right side of the SNS junction in Fig.~\ref{Fig20}(a). This model can happen naturally in an SOCSW with varying chemical potential, where the chemical potential varies from the topological regime to the nontopological regime resulting in the domain walls between the topological and nontopological regions~\cite{kell}. The ABSs can be found at the end of the topological region. For simplicity, here we consider a step jump in the chemical potential in going from the topologically nontrivial ($|\mu_0| < \sqrt{V_Z^2-\Delta_0^2}$) to the topologically trivial value ($|\mu_0| > \sqrt{V_Z^2-\Delta_0^2}$) keeping all the other parameters in these two regions to be the same. The ABS closest to zero energy in this model has energy oscillating with the chemical potential in the topological region as shown in Fig.~\ref{Fig20}(b), where the zero-energy ABS can be found at specific values of system parameters~\cite{Jay12}. 
In this section, we compare the conductance of an MZM with that of an ABS. We mention that the possible existence of ABS in the system can never be ruled out \textit{a priori}, and it is therefore important to take into account their possible effects on transport properties. In particular, we consider the ABS that may arise in the SOCSW model with a finite topological region and a semi-infinite nontopological region as shown in the right side of the SNS junction in Fig.~\ref{Fig20}(a). This model can happen naturally in an SOCSW with varying chemical potential where the chemical potential varies from the topological regime to the nontopological regime resulting in the domain walls between the topological and nontopological regions~\cite{kell}. The ABSs can be found at the end of the topological region. For simplicity, here we consider a step jump in the chemical potential in going from the topologically nontrivial ($|\mu_0| < \sqrt{V_Z^2-\Delta_0^2}$) to the topologically trivial value ($|\mu_0| > \sqrt{V_Z^2-\Delta_0^2}$) keeping all the other parameters in these two regions the same. The ABS closest to zero energy in this model has energy oscillating with the chemical potential in the topological region as shown in Fig.~\ref{Fig20}(b), where the zero-energy ABS can be found at specific values of system parameters~\cite{Jay12}. 

We consider this SOCSW in a junction with an $s$-wave superconducting lead. To calculate the conductance here, we first introduce a dissipation term $-i\Gamma\tau_0\otimes \sigma_0$ into the BdG Hamiltonian. The dissipation term is used to broaden the van Hove singularity of the BdG spectrum, so that we do not need to use a very fine energy grid in the numerical calculation. This dissipation term has been used previously to calculate conductance in topological NS junctions~\cite{Nag,Chunxiao}, though for different reasons. Our using a dissipation here could either be physically motivated as in Ref.~\cite{Chunxiao} or simply a technical artifice in handling the van Hove singularity. Figures~\ref{Fig20}(c)-(e) show the conductance of the SOCSW calculated for several chemical potential values in the topological region with all other parameters the same. The conductance for the zero-energy ABS may resemble the MZM tunneling conductance, i.e., it has a sharp rise at the voltage $e|V| = \Delta_s$ to a peak with a value near $G_M = (4-\pi)2e^2/h$ (see the inset in Fig.~\ref{Fig20}(d) or Ref.~\cite{seti2016}). One needs to be careful, therefore, in interpreting experimental data since accidental near-zero-energy ABS would produce tunneling conductance signatures quite similar to MZM themselves. For nonzero energy ABS, the ABS tunneling conductance peak shifts away from the threshold voltage $e|V| = \Delta_s$ (where $\Delta_s$ is the $s$-wave superconducting gap) toward a larger voltage value by the ABS energy normalized by the tunnel coupling between the lead and the system; see Figs.~\ref{Fig20}(c) and (e). 

\section{Conclusion}\label{sec6}
In this paper, we have calculated the zero-temperature dc current and conductance in various 1D voltage-biased SNS junctions involving topological and nontopological superconductors, considering both ideal spinful $p$-wave and realistic spin-orbit-coupled $s$-wave superconducting wires. For junctions with small transparencies, the presence of an MZM gives rise to a jump in the current and conductance at the gap-bias voltage $e|V| = \Delta_{\mathrm{lead}}$ where the superconducting gap edge is aligned with the MZM. If the superconducting lead has a BCS singularity at the gap edge then the tunneling conductance at the gap-bias voltage takes the value $G_M = (4-\pi)2e^2/h$ due to a single Andreev reflection from the MZM. However, this quantization no longer holds if the superconducting lead gap edge does not have the BCS singularity, e.g., $p$-wave superconductor or SOCSW with finite magnetic field.  For SNS junctions where both of the superconductors are topological (i.e., with one or two MZMs at each end), there is SGS in the $I$-$V$ curve or conductance profile due to MAR. However, for nontopological--topological superconductor junctions where the topological superconductor has only one MZM at each end, the SGS at small voltages is suppressed due to the mismatch in Andreev reflection spin-selectivity of the superconducting lead and the MZM. 

In contrast to the conventional SNS junction, where Cooper pairs are transferred across the junction with a charge of $2e$, for the topological SNS junction, the charge is transferred via the MZM in the units of $e$. As a result, for a perfectly transparent junction with an MZM at each end, the MZM contributes to a near zero-voltage current $I (V\rightarrow 0) = 2 e\Delta_{\mathrm{min}}/h$ where $\Delta_{\mathrm{min}}$ is the smallest gap in the junction. We note that this MZM near-zero voltage current is by no means universal or quantized because of the generic presence of the gap $\Delta_{\mathrm{min}}$ which surely varies from junction to junction. The same is also true for the case where there are two MZMs on one side and one MZM on the other side. This near zero-voltage dc current is half of the value for the conventional $s$-wave superconductor--normal--$s$-wave superconductor junction. However, for the case where there are two MZMs on both sides of the junction, the near zero-voltage current is $I (V\rightarrow 0) = 4 e\Delta_{\mathrm{min}}/h$ because each MZM can exchange a charge of $e$ between each other. For the case where there is a conventional $s$-wave superconductor on one side and one MZM on the other side of the junction, the current is zero because of the  difference in the Andreev-reflection spin selectivity of the $s$-wave superconductor and MZM, i.e., the $s$-wave superconductor allows only opposite-spin Andreev reflections and MZM favors equal-spin Andreev reflections. However, for the junction between a conventional $s$-wave superconductor and a Majorana Kramers pair the near-zero current for a perfect transparent junction is not zero but it is $I (V\rightarrow 0) = 4 e\Delta_{\mathrm{min}}/h$. This is due to the fact that the MZM pair can facilitate Andreev reflections in both spin channels.

We also calculated the conductance with an ABS in the SOCSW model arising from a finite topological and a semi-infinite nontopological region. For this junction, the energy of the ABS closest to zero energy oscillates with the chemical potential in the topological region. For the parameters where the ABS is at zero energy, the tunneling conductance may resemble that of Majorana, i.e., it has a step jump to a value $G_M$ at the gap-bias voltage $e|V| = \Delta_{\mathrm{lead}}$. However, when the energy of the ABS is nonzero, the conductance peak shifts away from the gap-bias voltage towards a larger voltage value by the ABS energy.

In conclusion, the tunneling conductance peaks for a conventional SNS junction occur at voltages $eV = \pm (\Delta^{\mathrm{nontopo}}_{\mathrm{L}}+ \Delta^{\mathrm{nontopo}}_{\mathrm{R}})$, where $\Delta^{\mathrm{nontopo}}_{\mathrm{L,R}}$ are the superconducting gaps of the left and right nontopological superconductors~\cite{octavio,Averin,Hurd,Bagwell,KBT}. For an SNS junction with an MZM at one side of the junction, the tunneling conductance peaks occur at voltages $eV = \pm \Delta_{\mathrm{nontopo}}$~\cite{Peng15,Denis,Yeyati,seti2016} and for an SNS junction with one MZM on both sides of the junction, the tunneling conductance develops peaks at $eV = \pm \Delta^{\mathrm{topo}}_{\mathrm{L}}$ and  $eV = \pm \Delta^{\mathrm{topo}}_{\mathrm{R}}$ where $\Delta^{\mathrm{topo}}_{\mathrm{L}}$ and $\Delta^{\mathrm{topo}}_{\mathrm{R}}$ are the superconducting gaps of the left and right topological superconductors~\cite{aguado,Badiane,Huang14}. For an SNS junction where both of the superconductors are identical SOCSWs, in the nontopological regime close to the topological phase transition, as the Zeeman field increases the zero-momentum gap ($\Delta = \sqrt{\mu_0^2 + \Delta_0^2} - V_Z$) shrinks and the tunneling conductance peaks move towards zero voltage with a rate $d(e|V_{\mathrm{tcp}}|)/dV_Z = -2$, where $V_{\mathrm{tcp}}$ is the voltage at which the tunneling conductance peak occurs. In the topological regime near the transition, as the zero-momentum gap ($\Delta = V_Z - \sqrt{\mu_0^2 + \Delta_0^2}$) reopen, the tunneling conductance peaks move towards larger voltage values with a rate $d(e|V_{\mathrm{tcp}}|)/dV_Z = 1$. This change in the dependence of the position of the tunneling conductance peaks with Zeeman field near the topological phase transition can serve as an evidence for the appearance of MZMs in the system.  

To this end, we would like to highlight the new finding of our paper. First, we find that the tunneling conductance of the MZM probed using a superconducting lead without a BCS singularity assumes a nonuniversal value, which decreases with decreasing junction transparency. We explicitly show this nonquantized conductance value for the case where the superconducting probe lead is either a topological or nontopological $p$-wave superconductor or SOCSW with finite magnetic field. Second, we also show that for the case where the superconducting probe lead is a $p$-wave superconductor with no topological channel, MAR are strongly suppressed due to fact that a nontopological $p$-wave superconductors is essentially an insulator with small Andreev reflection amplitudes. Third, we show that for the case where the superconducting probe lead is an $s$-wave superconductor and there is a Majorana Kramers pair in the topological superconductor, in the high transparency regime, the current and conductance near zero-voltage is not zero because there are two MZMs, which facilitate equal-spin Andreev reflections in \textit{two} different spin channels. This is in contrast to the case of an SNS junction between an $s$-wave superconductor and a topological superconductor with one MZM at the end. In this junction, MAR are strongly suppressed near zero voltage because of the difference between the Andreev-reflection spin selectivity of the $s$-wave superconductor and MZM.

Our theoretical results should serve as a definitive guide to future experiments on MZM using tunneling spectroscopy of topological SNS junctions.  We believe that such SNS experiments are now necessary since tunneling spectroscopy of NS junctions in nanowires has failed so far (in spite of $>$ 5 years of intense experimental activity) to manifest the predicted MZM quantization of zero-bias conductance although the zero-bias conductance peak itself seems to be observed generically.

\begin{acknowledgments}
We thank John Watson for stimulating discussions. This work is supported by Microsoft Station Q, LPS-MPO-CMTC, and JQI-NSF-PFC. J.D.S. acknowledges the funding from Sloan Research Fellowship and NSF-DMR-1555135 (CAREER). We acknowledge the University of Maryland supercomputing resources (http://www.it.umd.edu/hpcc) made available in conducting the research reported in this paper. 
\end{acknowledgments}

\appendix

\section{Remarks on Numerical Simulation}\label{appendix}
The scattering matrices at the left ($S_L$) and right NS interfaces ($S_R$) [Eq.~\eqref{eq:matchingeq}] can be calculated numerically from Kwant~\cite{kwant} by constructing the tight-binding models for the corresponding NS junctions. Since the scattering matrices given by Kwant are calculated using the current amplitudes with arbitrary phases at each energy, one can fix the phases by setting the largest element of the current amplitudes for every energy to be real. 

We note that Eqs.~\cref{subeq1} and ~\cref{subeq3} are invariant under the transformation:
\begin{align}
t_{L,R}^{\mathrm{in}}(E) \rightarrow t_{L,R}^{\mathrm{in}} (E) U_{L,R}^\dagger (E), \nonumber\\
\mathcal{J}_{L,R}^{\mathrm{in}}(E) \rightarrow U_{L,R}(E)\mathcal{J}_{L,R}^{\mathrm{in}}(E),
\end{align}
where $t_{L,R}^{\mathrm{in}}(E)$ are the transmission matrices at the left and right NS interfaces, $U_{L,R}(E)$ are unitary matrices, and $\mathcal{J}_{L,R}^{\mathrm{in}}(E)$ are the input current amplitudes from the left and right NS interfaces. By polar decomposition, there exists a unitary matrix $U_{L,R}(E)$ such that $t_{L,R}^{\mathrm{in}}(E) = \widetilde{t}_{L,R}^{\mathrm{in}}(E) U_{L,R}^\dagger (E)$, where
\beq\label{eq:transmatrix}
\widetilde{t}_{L,R}^{\mathrm{in}}(E) = \sqrt{t_{L,R}^{\mathrm{in}}(E)[t_{L,R}^{\mathrm{in}}(E)]^\dagger} = \sqrt{\mathds{1}-r_{L,R}(E) r_{L,R}^\dagger(E)}, 
\eeq
with $r_{L,R}$ being the reflection matrices at the left and right NS interfaces. For computational efficiency, we obtained only the reflection matrices $r_{L,R}$ from Kwant and used Eq.~\eqref{eq:transmatrix} to calculate the transmission matrix.

For the numerical evaluation of Eq.~\eqref{eq:curr}, we used an energy cutoff $E_c$ in the summation over energy where $E_c$ is chosen such that the calculation converges for each voltage $V$. The introduction of the energy cutoff sets the following constraint on the scattering matrix:
\begin{align}
S_{N}^e(E,E+eV) = S_{N}^h(-E,-(E+eV)) = -\mathds{1},
\end{align}
for all $E > E_c$. The above constraint is required for the unitarity of the scattering matrices to hold.

\end{document}